%% file: main.tex
\documentclass[preprint,prd,showpacs,superscriptaddress,preprintnumbers]{revtex4}

\def\LB{L\!B}

\usepackage{amsmath}
\usepackage{amssymb}
\usepackage{bm}
\usepackage{dcolumn}
\usepackage{graphicx}
\usepackage{subfigure}

\begin{document}

\date{\today}

\preprint{RIKEN-QHP-12}

\title{%
Tenth-Order QED Lepton Anomalous Magnetic Moment ---
Eighth-Order Vertices Containing a Second-Order Vacuum Polarization
}


\author{Tatsumi Aoyama}
\affiliation{Kobayashi-Maskawa Institute for the Origin of Particles and the Universe (KMI), Nagoya University, Nagoya 464-8602, Japan}
\affiliation{Nishina Center, RIKEN, Wako 351-0198, Japan }

\author{Masashi Hayakawa}
\affiliation{Department of Physics, Nagoya University, Nagoya 464-8602, Japan }
\affiliation{Nishina Center, RIKEN, Wako 351-0198, Japan }

\author{Toichiro Kinoshita}
\affiliation{Laboratory for Elementary-Particle Physics, Cornell University, Ithaca, New York 14853, USA }
\affiliation{Nishina Center, RIKEN, Wako 351-0198, Japan }

\author{Makiko Nio}
\affiliation{Nishina Center, RIKEN, Wako 351-0198, Japan }

\begin{abstract}
This paper reports the evaluation of
the tenth-order QED contribution to the lepton $g\!-\!2$ from
the gauge-invariant set of 2072 Feynman diagrams, called Set IV,
which are obtained by inserting a second-order lepton vacuum-polarization
loop into 518 eighth-order vertex diagrams
of four-photon exchange type. 
The numerical evaluation is carried out
by the adaptive-iterative Monte-Carlo integration routine {\sc vegas} 
using the {\sc fortran} codes written by the automatic code-generating algorithm
{\sc gencode}{\it N}.
Some of the numerical results are confirmed by comparison with
the values of corresponding integrals
that have been obtained previously by a different method.
The result for the mass-independent contribution of the Set IV
to the electron $g\!-\!2$ is $- 7.7296~ (48) (\alpha/\pi)^5$.
There is also a small mass-dependent contribution to the electron $g\!-\!2$
due to the muon loop: $- 0.01136~ (7) (\alpha/\pi)^5$.
The contribution of the tau-lepton loop is $-0.0000937~(104)~(\alpha/\pi)^5$. 
The sum of all these contributions  to the electron $g\!-\!2$ 
is $-7.7407~(49) (\alpha/\pi)^5$.
The same set of diagrams enables us to
evaluate the contributions to
the muon $g\!-\!2$ from the electron loop, muon loop, and tau-lepton loop.
They add up to $-46.95~ (17) (\alpha/\pi)^5$.

\end{abstract}

\maketitle

\section{Introduction}
\label{sec:intro}

The anomalous magnetic moment $g\!-\!2$ of the electron has played 
the central role in testing the validity of quantum electrodynamics (QED)
as well as the standard model. On the experimental side,
the latest measurement of $a_e \equiv (g\!-\!2)/2$ by the Harvard group has reached the precision
of $0.24 \times 10^{-9}$ 
\cite{Hanneke:2008tm,Hanneke:2010au}:
\begin{eqnarray}
a_e(\text{HV08})= 1~159~652~180.73~ (0.28) \times 10^{-12} ~~~[0.24 \text{ppb}]
~.
\label{a_eHV08}
\end{eqnarray}
The theoretical prediction thus far consists of 
QED corrections of up to the eighth order
\cite{Kinoshita:2005sm,Aoyama:2007dv,Aoyama:2007mn},
direct evaluation of hadronic corrections \cite{ Hagiwara:2011af, Davier:2010nc,
Krause:1996rf} \cite{Melnikov:2003xd,Bijnens:2007pz,Prades:2009tw,Nyffeler:2009tw} 
and electroweak corrections 
scaled down from their contributions to the muon $g\!-\!2$
\cite{Czarnecki:1995sz,Knecht:2002hr,Czarnecki:2002nt}. 
To compare the theory with the measurement 
(\ref{a_eHV08}),
we also need  the value of the fine structure constant $\alpha$
determined by a method independent of $g\!-2\!$ .
The best value of such an $\alpha$ available at present is one obtained 
from the measurement of $h/m_{\text{Rb}}$, the ratio of the Planck constant
and the mass of Rb atom,  
combined with the very precisely known Rydberg constant 
and $m_\text{Rb}/m_e$: \cite{Bouchendira:2010es}
\begin{eqnarray}
\alpha^{-1} (\text{Rb10}) = 137.035~999~037~(91)~~~[0.66 \text{ppb}].
\label{alinvRb10}
\end{eqnarray}  
With this  $\alpha$ the theoretical prediction of $a_e$ becomes 
\begin{eqnarray}
a_e(\text{theory}) = 1~159~652~181.13~(0.11)(0.37)(0.02)(0.77) \times 10^{-12},
\label{a_etheory}
\end{eqnarray}
where the first, second, third, and fourth uncertainties come
from the calculated eighth-order QED term \cite{Aoyama:2007mn},
a crude tenth-order estimate \cite{Mohr:2008fa},
the hadronic and electroweak contributions,
and the fine structure constant (\ref{alinvRb10}), respectively.
The theory (\ref{a_etheory})
is in good agreement with the experiment (\ref{a_eHV08}):  
\begin{eqnarray}
a_e(\text{HV08}) - a_e(\text{theory}) = -0.40~ (0.88) \times 10^{-12},
\end{eqnarray}
proving that QED (standard model) is in good shape even at this very high
precision.

An alternative and more sensitive test of QED is 
to calculate $\alpha$ from the experiment and theory of $g\!-2\!$ ,
both of which have very high precision, 
and compare it with $\alpha^{-1}$(Rb10).
The experiment and theory of the electron $g\!-\!2$ leads
\begin{eqnarray}
\alpha^{-1}(a_e 08) = 137.035~999~085~(12)(37)(2)(33)~~~[0.37 \text{ppb}],
\label{alinvae}
\end{eqnarray}
where the first, second, third, and fourth uncertainties come
from the eighth-order QED term, the tenth-order estimate, 
the hadronic and electroweak contributions,
and the measurement of $a_e$(HV08), respectively.

Although the uncertainty of
$\alpha^{-1} (a_e 08)$ in (\ref{alinvae}) is almost a factor 2 smaller than
that of $\alpha^{-1}$(Rb10), it is not a firm factor since it
 depends on the estimate of the tenth-order term,
which is only a crude guess \cite{Mohr:2008fa}.
For a more stringent test of QED, it is obviously necessary to evaluate
the actual value of the tenth-order term.
To meet this challenge we launched 
several years ago
a systematic program 
to evaluate the complete tenth-order term
\cite{Kinoshita:2004wi,Aoyama:2005kf,Aoyama:2007bs}.

The tenth-order QED contribution to the 
anomalous magnetic moment of an electron can be written as
\begin{equation}
	a_e^{(10)} 
	= \left ( \frac{\alpha}{\pi} \right )^5
         \left [ A_1^{(10)}
	+ A_2^{(10)} (m_e/m_\mu) 
	+ A_2^{(10)} (m_e/m_\tau) 
	+ A_3^{(10)} (m_e/m_\mu, m_e/m_\tau) \right ] ,
\label{eq:ae10th}
\end{equation}
where the electron-muon mass ratio $m_e/m_\mu $ is $ 4.836~331~66~(12) \times 10^{-3}$ and the electron-tau mass ratio $m_e/m_\tau $ is $ 2.875~64~(47) \times 10^{-4}$
\cite{Mohr:2008fa}.
In the rest of this article the factor $(\alpha/\pi)^5$ will be suppressed
for simplicity.

The contribution to the mass-independent term $A_1^{(10)}$ can be
classified into six gauge-invariant sets, further divided into
32 gauge-invariant subsets depending on the nature of closed
lepton loop subdiagrams.
Thus far, the numerical results of 29 gauge-invariant subsets, 
which consist of 3856 vertex diagrams, 
have been published \cite{Kinoshita:2005sm,Aoyama:2008gy,Aoyama:2008hz,Aoyama:2010yt,Aoyama:2010pk,Aoyama:2010zp,Aoyama:2011rm,Aoyama:2011zy}.
Five of these 29 subsets were also known analytically
\cite{Laporta:1994md,Aguilar:2008qj}.
They are in good agreement with our calculations.

In this paper we report the result of evaluation of $A_1^{(10)}$  from 
the set, called Set IV, which consists of 2072 Feynman diagrams.
Sec.~\ref{sec:construction} outlines our formulation
of Feynman-parametric integrals of Set IV.
Sec.~\ref{sec:residual} presents the residual renormalization formula,
which summarizes the result of derivation
described in detail in Appendix~\ref{sec:appendixa}.
Numerical results for several cases of mass dependence
are described in Secs.~\ref{sec:results}, \ref{sec:mass-dependent},
and~\ref{sec:results_muon}.
Sec.~\ref{sec:discussion} discusses the results obtained in this paper.

\section{Construction of Feynman-parametric Integrals}
\label{sec:construction}

All 2072 diagrams of Set IV can be derived from the 
518 eighth-order diagrams of four-photon-exchange type
\cite{Kinoshita:1981wm}, called Group V,
by inserting a second-order vacuum-polarization loop 
in the photon lines of Group V diagrams  in all possible ways.
In practice, we have therefore to deal with only 518 diagrams.
\begin{figure}[Ht]
\includegraphics[width=16cm]{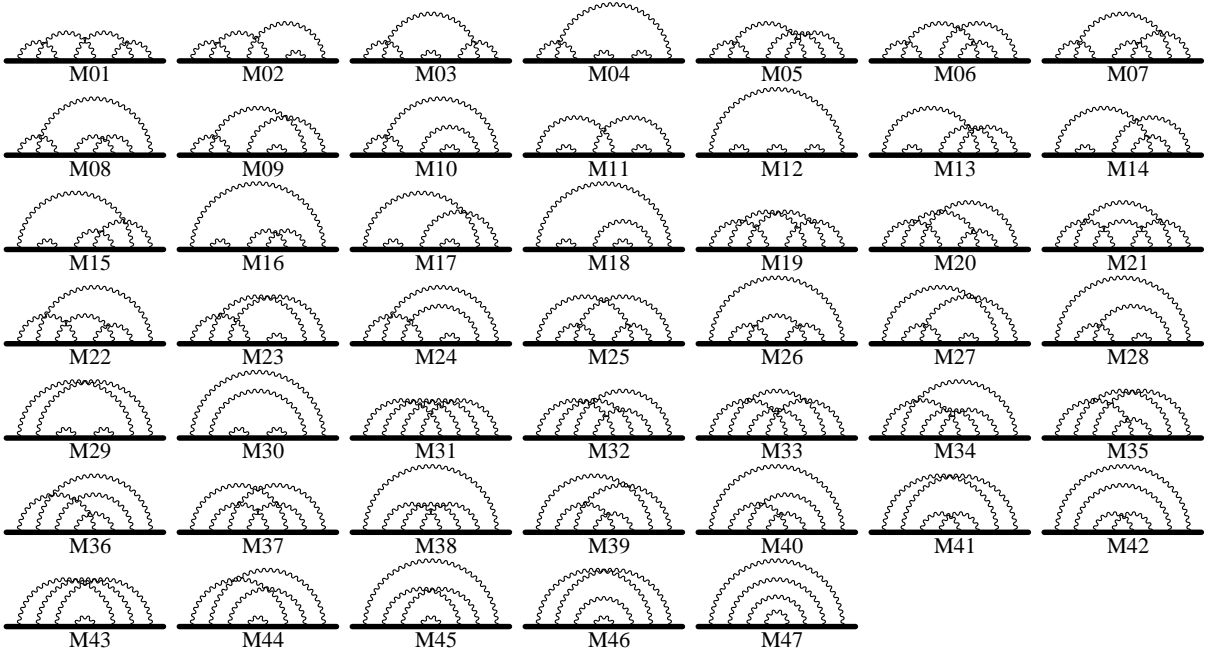}
\caption{
\label{fig:M8} 
The eighth-order Group V diagrams. The solid line represents the electron 
propagating in a  weak constant magnetic field. 
}
\end{figure}
This can be reduced further to the 47 self-energy-like diagrams
of Fig.~\ref{fig:M8} as follows. 

Let $\Lambda^\nu$ be the sum of 7 vertex diagrams
that are obtained from any self-energy-like diagram $\Sigma (p)$ 
of Fig.~\ref{fig:M8} 
by inserting a magnetic vertex $\gamma^\nu$ in all possible ways.
The set of these vertex diagrams,
taking account of doubling due to time-reversal, 
 represents the original 518 vertex diagrams.
The next step is to rewrite this $\Lambda^\nu$ as
\begin{equation}
\Lambda^\nu (p, q) \simeq - q^\mu \left [ \frac{\partial \Lambda_\mu (p, q)}
{\partial q_\nu} \right ]_{q=0} - \frac{\partial \Sigma (p)}{\partial p_\nu}
\label{WTid}
\end{equation}
for small $q$, with the help of the Ward-Takahashi(WT) identity,
where $p-q/2$ and $p+q/2$ are the 4-momenta of incoming and outgoing
lepton lines and $(p-q/2)^2 =(p+q/2)^2 = m^2$.
The $g\!-\!2$ term is projected out from the right-hand side of 
Eq.~(\ref{WTid}).

The properties of the Feynman-parametric integrals corresponding to 
the diagrams of Fig.~\ref{fig:M8}
have been studied and described in detail in Ref.~\cite{Aoyama:2007mn}.  
Each diagram ${\cal G}$ of Fig.~\ref{fig:M8} is represented by
a momentum integral using the Feynman-Dyson rule.
Introducing Feynman parameters $z_1, z_2, \ldots, z_7$
for the electron propagators and $z_a, z_b, z_c, z_d$
for the photon propagators,
we carry out the momentum integration analytically
using a home-made program written in FORM \cite{Vermaseren:2000nd}, 
which gives
an integral of the form
\begin{equation}
M_{\cal G} =  \left (\frac{-1}{4} \right )^4 3! \int (dz)_{\cal G}
 \left[\frac{1}{3} \left ( \frac{E_0 + C_0}{U^2V^3} 
                        +  \frac{E_1+C_1}{U^3V^2} + \cdots \right) 
                 + \left (\frac{N_0+Z_0}{U^2 V^4} 
                       +  \frac{N_1+Z_1}{U^3 V^3}+\cdots \right) \right],
\label{M10}
\end{equation}
where ${E_n}, {C_n}, {N_n}$ and ${Z_n}$ 
are functions of Feynman parameters.
The subscript $n$ of ${E_n}$, etc., indicates that
it is the $n$ contraction terms of diagonalized loop momenta and
proportional to the product of $n$ factors of $B_{ij}$'s.
The ``symbolic'' building blocks $A_i, B_{ij}, C_{ij}$,
for $i, j = 1, 2, \ldots, 7$ are also functions of Feynman parameters.
$U$ is the Jacobian of transformation from the momentum space variables to
Feynman parameters. 
$V$ is obtained by combining denominators of all propagators into one
with the help of Feynman parameters.
It has a form common to all diagrams of Fig.~\ref{fig:M8}:
\begin{equation}
V= \sum_{i=1}^7 z_i (1 - A_i) m_i^2 + \sum_{k=a}^d z_k \lambda_k^2,
\label{defv}
\end{equation}
where $m_i$ and $\lambda_k$ are the rest masses of electron $i$
and photon $k$, respectively.
$A_i$ is the {\it scalar current} defined by
\begin{equation}
A_i = \delta_{ij} - \frac{1}{U} \sum_{j=1}^7 z_j B_{ij},
\end{equation}
and 
\begin{equation}
(dz)_{\cal G} = \prod_{i \in {\cal G}} dz_i \delta (1- \sum_{i \in {\cal G}} z_i ).
\end{equation}
See, for example, Ref.~\cite{Kinoshita:1990} for definitions of $B_{ij}$ and $C_{ij}$.
The form of $A_i$ as a function of Feynman parameters
depends on the structure of individual diagram. However, 
as is shown in Eq.~(\ref{defv}), the expression of $V$
in terms of $A_i$ is identical for all diagrams of Fig.~\ref{fig:M8}.
Individual diagram of Fig.~\ref{fig:M8} will be denoted
as $M_{\cal G}$ and their assembly
will be collectively denoted as $M_8$.

\begin{figure}[Ht]
\includegraphics[width=16cm]{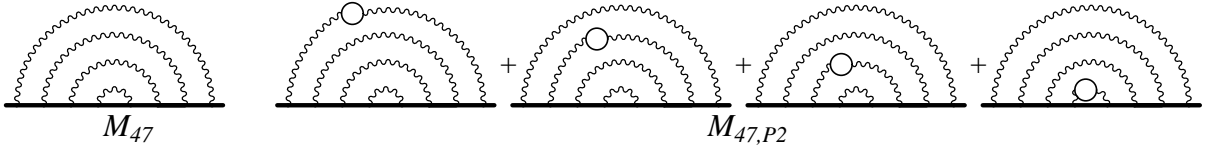}
\caption{
\label{fig:M47P2} 
The eighth-order diagram $M_{47}$ of Group V and 
the tenth-order diagram $M_{47,P2}$ of Set IV.
The diagram $M_{47,P2}$ represents the sum of diagrams obtained by inserting  
a second-order vacuum-polarization 
loop into each  of four photon lines of the eighth-order diagram $M_{47}$.
}
\end{figure}

We have developed two independent sets of numerical programs of $M_{\cal G}$ based on the WT-summed amplitudes. The first formulation was developed in
1970's and given in Ref.~\cite{Kinoshita:1981vs}.    The second formulation
used the automation code {\sc gencode}{\it N} \cite{Aoyama:2005kf,Aoyama:2007bs}. The unrenormalized amplitudes and the UV-subtraction terms
are the same, but the IR-subtractions are slightly
different in two formulations.  The detail of UV- and IR-subtraction terms in 
the second formulation is briefly described in Sec.~\ref{sec:residual}. 
After taking account of the difference in two formulations, 
the equivalence of two formulations is established \cite{Aoyama:2007bs}.
Once we have the correct programs of the eighth-order Group V diagrams,
the insertion of a vacuum-polarization loop is an easy task to carry out.
Fig.~\ref{fig:M47P2} shows a typical self-energy-like diagram of the tenth-order
Set IV. 

As is well-known, the insertion of a vacuum-polarization loop in
an internal photon line can be expressed 
as a superposition of massive vector particle propagators.
In other words all we have to do is to replace the mass square $\lambda^2$ of 
one of the photons in Eq.~(\ref{defv})
by $p(t)$:
\begin{equation}
\lambda^2 \longrightarrow  p(t) \equiv  \frac{4 m_{vp}^2}{1 - t^2},
\label{vector}
\end{equation}
where $m_{vp}$ is the mass of the fermion forming the vacuum-polarization 
loop, 
to multiply the resulting eighth-order integral with the spectral function
\begin{equation}
\rho_2 (t) = \frac{t^2}{1-t^2} \left ( 1- \frac{1}{3} t^2 \right ),
\label{spectral}
\end{equation}
and to integrate over the interval $0 \leq t < 1$.

This is easy to implement in the second formulation 
\cite{Aoyama:2005kf,Aoyama:2007bs} 
since the function $V$ is unambiguously identifiable.
Unfortunately, in the first formulation \cite{Kinoshita:1981vs}, 
it is difficult to implement this procedure for some diagrams
because the ``{\it denominator function} $V$'' 
was used to replace parts 
of numerators in order to reduce the size of
integrands and accelerate the computing speed.
For this reason, it is difficult to  apply Eqs.~(\ref{vector})
and (\ref{spectral}) to these integrals.
Thus, direct comparison of two methods is 
feasible only for those of
Set IV diagrams in which the function $V$ can be clearly distinguished
from other terms of the numerator. 
We therefore report here only the results of the second formulation.
Since the equivalence of two methods has been well established
\cite{Aoyama:2007mn},
this does not diminish the reliability of our numerical results.


\section{Residual renormalization}
\label{sec:residual}

In our approach based on numerical integration the integrals 
of individual diagrams must be made convergent before they
are integrated numerically.
This is achieved in the following manner. 

Suppose the integral $M_{\cal G}$ has a UV divergence arising from
a subdiagram ${\cal S}$.
Then we construct another integral $K_{\cal S} M_{\cal G}$
by applying a {\it K}-operation,
which identifies and extracts the UV divergent part
 of $M_{\cal G}$ by a simple power counting rule.
This integral has the following properties:
\begin{itemize}

\item  It has the same domain of integration 
and the same UV divergence as $M_{\cal G}$.
Thus it subtracts the
UV divergence of the latter {\it point-by-point} in the domain of integration.

\item  If ${\cal S}$ is a vertex diagram,
 {\it K}-operation $\mathbb{K}_{\cal S}$ on $M_{\cal G}$
factorizes exactly into the product of lower-order quantities as
\begin{equation}
\mathbb{K}_{\cal S} M_{\cal G} = L_{\cal S}^{\rm UV} M_{{\cal G}/{\cal S}}.
\label{KSMG1}
\end{equation}
If ${\cal S}$ is a self-energy diagram,
 {\it K}-operation $\mathbb{K}_{\cal S}$ on $M_{\cal G}$
turns exactly into the sum of two terms of the form
\begin{equation}
\mathbb{K}_{\cal S} M_{\cal G} = {\delta m}_{\cal S}^{\rm UV} M_{{\cal G}/{{\cal S}(i^*)}}
            + B_{\cal S}^{\rm UV} M_{{\cal G}/{{\cal S}(i^{'})}}.
\label{KSMG2}
\end{equation}
Here $L_{\cal S}^{\rm UV}$, $B_{\cal S}^{\rm UV}$, and ${\delta m}_{\cal S}^{\rm UV}$
are UV-divergent parts of the renormalization constants $L_{\cal S}$,
$B_{\cal S}$, and ${\delta m}_{\cal S}$.
$M_{{\cal G}/{\cal S}}$ is the magnetic moment
corresponding to the diagram ${\cal G}/{\cal S}$ obtained by shrinking
the subdiagram ${\cal S}$ of ${\cal G}$ to a point.
See Ref.~\cite{Kinoshita:1990} for further details.
\end{itemize}

An IR divergence of $M_{\cal G}$ arises from a subdiagram ${\cal T}$
that is the reduced diagram ${\cal T} \equiv {\cal G}/{\cal S}$ of a self-energy 
subdiagram ${\cal S}$ of the diagram ${\cal G}$.
In this case we run into two kinds of IR divergence.
One arises when a self-energy subdiagram ${\cal S}$ behaves as a self-mass term.
The standard mass-renormalization on ${\cal G}$ subtracts
${\delta m}_{\cal S} M_{{\cal G}/{{\cal S}(i^*)}}$ from $M_{\cal G}$ while
$K_{\cal S}$ operation of Eq.~(\ref{KSMG2}) subtracts
${\delta m}_{\cal S}^{\rm UV} M_{{\cal G}/{{\cal S}(i^*)}}$.
Thus, after subtraction by the $K_{\cal S}$ operation, we are left with
$({\delta m}_{\cal S} - {\delta m}_{\cal S}^{\rm UV}) M_{{\cal G}/{{\cal S}(i^*)}}$,
which has a linear IR divergence because of divergent
$M_{{\cal G}/{{\cal S}(i^*)}}$, except when
${\delta m}_{\cal S} = {\delta m}_{\cal S}^{\rm UV}$.
The easiest way to deal with this problem is to subtract ${\delta m}_{\cal S}$
entirely instead of only ${\delta m}_{\cal S}^{\rm UV}$.
We call this {\it R}-subtraction, which is incorporated in {\sc gencode}{\it N}.

The other IR divergence  occurs when a self-energy-like subdiagram ${\cal S}$
behaves as a magnetic moment amplitude. The remaining diagram ${\cal T}$ can be
mimicked by a vertex diagram by shrinking the subdiagram ${\cal S}$ to a point. 
This divergence is only logarithmic and the subtraction term
can be constructed by applying the {\it I}-subtraction $\mathbb{I}_{\cal T}$
on the UV-finite 
amplitude $\underline{M}_{\cal G}$, which is shown to factorize as
\cite{Aoyama:2007bs}
\begin{equation}
\mathbb{I}_{\cal T} \underline{M}_{\cal G} = L_{\cal T}^{\rm R} \underline{M}_{\cal S},
\label{ITMG1}
\end{equation}
where $L_{\cal T}^{\rm R}$ is the part of the vertex 
renormalization constant $L_{\cal T}$ that remains after all UV-divergent pieces
are subtracted out.

These operations, carried out for all divergent subdiagrams
of the unrenormalized integral $M_{\cal G}$, create
a UV-finite and IR-finite integral $\Delta M_{\cal G}$.
For a full account of these operations see 
Refs.~\cite{Aoyama:2005kf,Aoyama:2007bs}.

Since this scheme is different from the standard on-the-mass-shell renormalization,
it is necessary to make an adjustment, called residual renormalization,
which accounts for  the difference of the standard renormalization 
and the UV-divergent (and IR-divergent)  parts generated by {\it K}-operation
(and {\it I/R}-subtractions).

The residual renormalization terms of individual diagrams 
must then be summed up over all diagrams involved.
As the order of perturbation increases the total number of terms
contributing to the residual renormalization increases rapidly
so that it will become harder and harder to manage.
Fortunately, the sum of all residual terms can be expressed
concisely in terms of magnetic moments and finite parts of
renormalization constants of lower orders,
whose structure is closely related to that 
of the standard on-the-mass-shell renormalization.
This observation enables us to obtain the sum of residual
renormalization terms of all integrals starting from the expression of
the standard renormalization.
This approach is described in detail in Appendix~\ref{sec:appendixa} 
for the eighth-order $g\!-\!2$
after simpler cases of fourth- and  sixth-orders are 
described for illustration of our method.

Since diagrams of Set IV are obtained 
from the magnetic moment contribution $M_8$ of
518 eighth-order vertices of four-photon-exchange
type by inserting a second-order
vacuum-polarization subdiagram in all possible ways,
the residual renormalization term of the Set IV is readily derived
from that of the residual renormalization term of $M_8$.
Namely, insertion of a closed loop of the lepton $l_2$ in the internal photon
lines of Group V diagrams of lepton $l_1$ given in Eq.~(\ref{a_8}) in all possible ways leads to the renormalized contribution of Set IV to the lepton
$g\!-\!2$ of the form:
\begin{align}
	A^{(10)} [\text{Set~IV}^{(l_1l_2)}] &= \Delta M_{8,P2}^{(l_1l_2)}
\nonumber \\
	&  - 5 \Delta M_{6,P2}^{(l_1l_2)} \Delta \LB_2
\nonumber \\
	&  - 5 \Delta M_6 \Delta \LB_{2,P2}^{(l_1l_2)}
\nonumber \\
	& + \Delta M_{4,P2}^{(l_1l_2)} (-3 \Delta \LB_{4} +9 (\Delta \LB_{2})^2 )
\nonumber \\
	& + \Delta M_{4} (-3 \Delta \LB_{4,P2}^{(l_1l_2)} +18 \Delta \LB_{2} \Delta \LB_{2,P2}^{(l_1l_2)} )
\nonumber \\
	& + M_{2,P2}^{(l_1l_2)} (-\Delta \LB_{6} +6 \Delta \LB_4 \Delta \LB_2  -5 (\Delta \LB_2)^3) 
\nonumber \\
	& + M_2 (-\Delta \LB_{6,P2}^{(l_1l_2)} +6 \Delta \LB_{4,P2}^{(l_1l_2)} \Delta \LB_2
\nonumber \\
           &~~~~+6 \Delta \LB_4 \Delta \LB_{2,P2}^{(l_1l_2)} -15 (\Delta \LB_2)^2 \Delta \LB_{2,P2}^{(l_1l_2)}) 
\nonumber \\
	& + M_{2,P2}^{(l_1l_2)} \Delta {\delta m}_{4} (4 \Delta L_{2^*}+\Delta B_{2^*}) 
\nonumber \\
	& + M_2 \Delta {\delta m}_{4,P2}^{(l_1l_2)} (4 \Delta L_{2^*}+\Delta B_{2^*})
\nonumber \\
	& + M_2 \Delta {\delta m}_{4} (4 \Delta L_{2^*,P2}^{(l_1l_2)}+\Delta B_{2^*,P2}^{(l_1l_2)}), 
	\label{a8_setIV}
\end{align}
where superscripts such as $(l_1l_1)$
and $(l_2l_2)$ are omitted for terms which are independent
of rest mass.
See Appendix~\ref{sec:appendixa} for the explanation of notations.

$\Delta M_{8,P2}^{(l_1l_2)}$ is the sum of 74 WT-summed integrals
enhanced by the insertion of vacuum-polarization-loop.
Each of these 74 integrals is finite by our construction.
Individual terms of residual renormalization are also UV- and IR-finite
by construction.
Eq.~(\ref{a8_setIV}) thus maintains that 
$A^{(10)} [\text{Set~IV}^{(l_1l_2)}]$, which represents the quantity
renormalized in the standard manner,
can be expressed as the sum of completely finite quantities,
each of which can thus be integrated by numerical means.

We should like to emphasize that
Eq.~(\ref{a8_setIV}) is analytically exact and involves no approximation
as far as the subtraction term factorizes exactly as 
in Eqs.~(\ref{KSMG1}), (\ref{KSMG2}), and~(\ref{ITMG1}).


\section{Numerical evaluation of $A_1^{(10)} [\text{Set IV} ]$}
\label{sec:results}

\input tableEe.tex

\input res-tableEe.tex

$\Delta M_{\alpha,P2}$,
which is made UV-finite by {\it K}-operation and IR-finite by {\it I/R}-subtractions,
is integrated numerically by the adaptive Monte-Carlo integration
routine {\sc vegas} \cite{Lepage:1977sw}.
The result for $(l_1l_2)=(ee)$ are
listed in Tables~\ref{table:setIV_ee_1} and \ref{table:setIV_ee_2}.
Auxiliary quantities needed for carrying out the residual renormalization
are listed in Table~\ref{Table:residual_const}.
Notations  are those of Eq.~(\ref{a8_setIV}).
Substituting these quantities in Eq.~(\ref{a8_setIV}) we obtain
\begin{equation}
A_1^{(10)} [\text{Set~IV}^{(ee)}] = -7.729~6~(48).
\label{set4ee}
\end{equation}
%


\section{Numerical evaluation of
$A_2^{(10)} (m_e/m_\mu)$ and
$A_2^{(10)} (m_e/m_\tau)$}
\label{sec:mass-dependent}

\input tableEm.tex

\input res-tableMe.tex

Once {\sc fortran} programs for mass-independent contributions are obtained,
it is straightforward to evaluate the contribution of mass-dependent
terms such as $A_2^{(10)} (m_e/m_\mu)$.
We simply have to choose an appropriate rest mass for the loop fermion $l_2$.
The result for $A_2^{(10)} (m_e/m_\mu)$ is listed in
Tables~\ref{table:setIV_em_1} and
\ref{table:setIV_em_2}.
From these Tables and the additional data listed in
Table~\ref{Table:residual_const_mass_dep} we obtain
\begin{equation}
A_2^{(10)} [\text{Set~IV}^{(em)}] = - 0.011~36~(7). 
\label{set4em}
\end{equation}
We have also computed the contribution of tau-particle loop
$A_2^{(10)} (m_e/m_\tau)$, which we give without details:
\begin{equation}
A_2^{(10)} [\text{Set~IV}^{(et)}] = - 0.000~093~7~(104) .
\label{set4et}
\end{equation}

The contribution of the muon loop (\ref{set4em}) is about 0.13 \%
of the electron loop contribution (\ref{set4ee}),
while the contribution of the tau-lepton loop (\ref{set4et})
is much smaller than the uncertainty of (\ref{set4ee})
and hence completely negligible at present.
%


\section{Contribution to the muon $g\!-\!2$}
\label{sec:results_muon}

\input tableMe.tex
\input tableMt.tex

The muon $g\!-\!2$ also receives contributions from the Set IV.
The contributions coming from the electron loop 
$(l_1l_2)=(me)$ are listed in Tables~\ref{table:setIV_me_1} and \ref{table:setIV_me_2}.
Auxiliary quantities needed to carry out the residual renormalization
are listed in Table~\ref{Table:residual_const_mass_dep}.
From these quantities we obtain
\begin{equation}
A_2^{(10)} [\text{Set~IV}^{(me)}] = - 38.79~(17).
\label{set4me}
\end{equation}
We also obtained the contribution of the tau-lepton loop
$A_2^{(10)} (m_e/m_\tau)$.  The result is listed in
Tables~\ref{table:setIV_mt_1} and
\ref{table:setIV_mt_2}.
From these Tables and the additional data listed in 
Table~\ref{Table:residual_const_mass_dep} we obtain
\begin{equation}
A_2^{(10)} [\text{Set~IV}^{(mt)}]  = - 0.435~7~(25).
\label{set4mt}
\end{equation}
Including the mass-independent contribution (\ref{set4ee}),
the total contribution to the muon $g\!-\!2$ amounts to
\begin{equation}
A_2^{(10)} [\text{Set~IV}^{(me+mm+mt)}]  = - 46.95~(17).
\label{set4mug-2_ldg}
\end{equation}


\section{Discussion}
\label{sec:discussion}

Since the reliability of the eighth-order term $M_8$ is crucial for 
the validity of our work on the Set IV,
let us sketch briefly how we established the  validity of $M_8$.
See Ref.~\cite{Aoyama:2007mn} for detailed accounts.
Our approach was to evaluate the
diagrams contributing to $M_8$ in two independent ways.
The first method is to apply
the scheme formulated more than 30 years ago \cite{Kinoshita:1981vs}.
The revised numerical evaluation by this formulation
 was reported recently \cite{Aoyama:2007dv,Aoyama:2007mn}.
The second approach relies on the {\sc fortran} codes written
by the automatic code-generator {\sc gencode}{\it N}
\cite{Aoyama:2005kf,Aoyama:2007bs}.
This method treats the 
self-mass renormalization terms and IR divergent terms differently 
from the first method so that
they can be regarded as practically independent of each other.
Comparison of the results of these two methods revealed that the first one had 
a subtle inconsistency in the handling of some IR subtraction terms.
Correcting this error we now have two independent evaluations
of $M_8$ which agree with each other within
the precision of numerical integration \cite{Aoyama:2007mn}.

Although we have not shown the analytic equivalence of the two methods directly,
we are fully convinced that they are indeed equivalent
by proving that they agree to 13 or 14 digits
(in double precision) at all arbitrarily chosen points in the domain
of integration.
Only last few digits disagree due to difference in rounding off.

The validity of integrals of Set IV relies on the fact that
two versions of $M_8$ agree completely with each other.
As was noted in Sec.~\ref{sec:construction}
we actually used only the second version of $M_8$ to build
integrals of tenth-order diagrams of Set IV, because of
a technical problem in the first version.  However,
we are convinced that the integrals of Set IV are indeed bug-free.

As is seen from (\ref{set4me}) the contribution of Set IV to
the muon $g\!-\!2$ is sizable,
which is not unexpected.  This is because 
the order of magnitude
of the contribution of the dominant $(me)$ term can be readily estimated,
noting that the leading $\ln (m_\mu/m_e)$ term is determined
by the charge renormalization procedure.  This leads to
\begin{equation}
A_2^{(10)} [\text{Set~IV}^{(me)}]  \sim 4 K_2 a_e^{(8)}[\text{Group~V}] \sim 
-31.0
\label{set4mug-2}
\end{equation}
where the factor 4 comes from the number of virtual photon lines
of $a_e^{(8)}[\text{Group~V}]$ into which a vacuum-polarization loop can be inserted,
$a_e^{(8)}[\text{Group~V}] \simeq -2.179~(3)$
\cite{Aoyama:2007dv,Aoyama:2007mn}, and the enhancement factor \cite{Kinoshita:2005sm}
\begin{equation}
K_2 \sim \frac{2}{3} \ln (m_\mu/m_e) \sim  3.6~.
\label{K2}
\end{equation}
The value (\ref{set4mug-2}) may be regarded as
a fair approximation to (\ref{set4me}).

By now we have evaluated the complete set of tenth-order diagrams containing vacuum-polarization subdiagrams
\cite{Kinoshita:2005sm,Aoyama:2008gy,Aoyama:2008hz,Aoyama:2010yt,Aoyama:2010pk,
Aoyama:2010zp,Aoyama:2011rm,Aoyama:2011zy}.
(Note that the remaining Sets have no vacuum-polarization loop.)
In particular its $(me)$ contribution to the muon $g\!-\!2$, namely 
all sets excluding light-by-light scattering loops, is given by
\begin{equation}
A_2^{(10)}[\text{All sets excluding l-l loops}]^{(me)} \simeq 48.88~(19).
\label{Xsum_noll}
\end{equation}
This may be compared with the corresponding result $\Delta_{(10)}^{(I)} \simeq 32$ obtained by an  estimate based on the renormalization group method
\cite{Kataev:2006yh}.


\begin{acknowledgments}
This work is supported in part by the JSPS Grant-in-Aid for
Scientific Research (C)19540322, (C)23540331, and (C)20540261.
T. K.'s work is supported in part by the U. S. National Science Foundation
under Grant NSF-PHY-0757868, and the International Exchange Support
Grants (FY2010) of RIKEN.
T. K. thanks RIKEN for the hospitality extended to him
while a part of this work was carried out.
Numerical calculations are conducted on the
RIKEN Super Combined Cluster System (RSCC) and the
RIKEN Integrated Cluster of Clusters (RICC) supercomputing systems.
Special thanks are due to  late~Dr.~T.~Shigetani and High Performance
Computing group  of RIKEN's Advanced Center for Computing and Communication.

\end{acknowledgments}


\appendix

\section{Summing up Residual Renormalization Terms}
\label{sec:appendixa}

\input appA.tex

\bibliographystyle{apsrev}
\bibliography{b}

\end{document}

%% file: tableEe.tex
\begingroup
\renewcommand{\baselinestretch}{1.0}

\begin{table*}
  \caption{%
Contributions of diagrams $M_{01,P2}$,\dots, $M_{24,P2}$ of Set~IV to $a_e$ 
for $(l_1l_2) = (ee)$. 
The multiplicity $n_F$ is the number of vertex diagrams 
represented by the integral and 
is incorporated in the numerical value. 
First 50 iterations are carried out using $1 \times 10^{8}$ sampling
points per iteration. The integrations are continued with  
$1 \times 10^{9}$ sampling points per iteration and iterated as given in  
the second number of the fifth column. 
The integrals $M_{12,P2}$, $M_{16,P2}$, and $M_{18,P2}$ are evaluated with quadruple
precision.
All other integrals are evaluated with double precision.
  \label{table:setIV_ee_1}
}

\begin{ruledtabular}
    \begin{tabular}{lcdcc}
\multicolumn{1}{c}{Integral} & 
\multicolumn{1}{c}{$n_F$} &
\multicolumn{1}{c}{\hspace*{4em}\parbox[t]{10em}{Value (Error) \\[-1ex] including $n_F$}} &
\multicolumn{1}{c}{\parbox[t]{10em}{Sampling per \\[-1ex]  iteration}} &
\multicolumn{1}{c}{\parbox[t]{4em}{No. of \\[-1ex] iterations}} \\[3ex]
\hline

$M_{01,P2}$ & 28 &   -0.509~62~( 38) & $ 1\times 10^8,~~1\times 10^9 $ & 50,~~180\\
$M_{02,P2}$ & 56 &    0.060~41~( 98) & $ 1\times 10^8,~~1\times 10^9 $ & 50,~~200\\
$M_{03,P2}$ & 28 &    0.829~28~( 60) & $ 1\times 10^8,~~1\times 10^9 $ & 50,~~200\\
$M_{04,P2}$ & 56 &    1.497~37~(126) & $ 1\times 10^8,~~1\times 10^9 $ & 50,~~360\\
$M_{05,P2}$ & 56 &    0.130~36~( 46) & $ 1\times 10^8,~~1\times 10^9 $ & 50,~~180\\
$M_{06,P2}$ & 56 &   -1.084~60~( 94) & $ 1\times 10^8,~~1\times 10^9 $ & 50,~~220\\
$M_{07,P2}$ & 56 &   -1.178~02~( 48) & $ 1\times 10^8,~~1\times 10^9 $ & 50,~~180\\
$M_{08,P2}$ & 56 &   -1.415~81~(121) & $ 1\times 10^8,~~1\times 10^9 $ & 50,~~360\\
$M_{09,P2}$ & 56 &   -0.011~71~( 95) & $ 1\times 10^8,~~1\times 10^9 $ & 50,~~360\\
$M_{10,P2}$ & 56 &   -0.816~12~(107) & $ 1\times 10^8,~~1\times 10^9 $ & 50,~~360\\
$M_{11,P2}$ & 28 &    0.768~73~( 48) & $ 1\times 10^8,~~1\times 10^9 $ & 50,~~200\\
$M_{12,P2}$ & 28 &   -1.631~37~( 37) & $ 1\times 10^8,~~1\times 10^9 $ & 50,~~  5\\
$M_{13,P2}$ & 56 &   -2.353~59~( 45) & $ 1\times 10^8,~~1\times 10^9 $ & 50,~~200\\
$M_{14,P2}$ & 56 &    0.685~64~( 83) & $ 1\times 10^8,~~1\times 10^9 $ & 50,~~200\\
$M_{15,P2}$ & 56 &    0.461~55~( 43) & $ 1\times 10^8,~~1\times 10^9 $ & 50,~~200\\
$M_{16,P2}$ & 56 &    1.763~95~(111) & $ 1\times 10^8,~~1\times 10^9 $ & 50,~~025\\
$M_{17,P2}$ & 56 &    3.290~90~(120) & $ 1\times 10^8,~~1\times 10^9 $ & 50,~~340\\
$M_{18,P2}$ & 56 &   -0.052~73~( 53) & $ 1\times 10^8,~~1\times 10^9 $ & 50,~~  5\\
$M_{19,P2}$ & 28 &   -1.403~73~(  6) & $ 1\times 10^8,~~1\times 10^9 $ & 50,~~180\\
$M_{20,P2}$ & 56 &    0.856~32~( 43) & $ 1\times 10^8,~~1\times 10^9 $ & 50,~~180\\
$M_{21,P2}$ & 28 &    0.360~89~(  5) & $ 1\times 10^8,~~1\times 10^9 $ & 50,~~180\\
$M_{22,P2}$ & 56 &   -0.743~60~( 46) & $ 1\times 10^8,~~1\times 10^9 $ & 50,~~180\\
$M_{23,P2}$ & 56 &   -1.120~08~( 90) & $ 1\times 10^8,~~1\times 10^9 $ & 50,~~200\\
$M_{24,P2}$ & 56 &    0.870~63~( 59) & $ 1\times 10^8,~~1\times 10^9 $ & 50,~~200\\

  \end{tabular}
  \end{ruledtabular}
\end{table*}

\endgroup

\begingroup
\renewcommand{\baselinestretch}{1.0}
\begin{table*}
  \caption{%
Contributions of diagrams $M_{25,P2}$,\dots, $M_{47,P2}$ of Set~IV to $a_e$ 
for $(l_1l_2) = (ee)$. 
The multiplicity $n_F$ is the number of vertex diagrams 
represented by the integral and 
is incorporated in the numerical value. 
All integrals are evaluated with double precision.
  \label{table:setIV_ee_2}
}

\begin{ruledtabular}
    \begin{tabular}{lcdcc}
\multicolumn{1}{c}{Integral} & 
\multicolumn{1}{c}{$n_F$} &
\multicolumn{1}{c}{\hspace*{4em}\parbox[t]{10em}{Value (Error) \\[-1ex] including $n_F$}} &
\multicolumn{1}{c}{\parbox[t]{10em}{Sampling per \\[-1ex]  iteration}} &
\multicolumn{1}{c}{\parbox[t]{4em}{No. of \\[-1ex] iterations}} \\[3ex]
\hline

$M_{25,P2}$ & 28 &   -0.696~52~( 25) & $ 1\times 10^8,~~1\times 10^9 $ & 50,~~180\\
$M_{26,P2}$ & 28 &   -0.432~10~( 51) & $ 1\times 10^8,~~1\times 10^9 $ & 50,~~200\\
$M_{27,P2}$ & 56 &    1.120~35~( 87) & $ 1\times 10^8,~~1\times 10^9 $ & 50,~~200\\
$M_{28,P2}$ & 56 &    0.783~12~( 94) & $ 1\times 10^8,~~1\times 10^9 $ & 50,~~240\\
$M_{29,P2}$ & 28 &    1.495~00~( 92) & $ 1\times 10^8,~~1\times 10^9 $ & 50,~~200\\
$M_{30,P2}$ & 28 &   -0.850~74~( 95) & $ 1\times 10^8,~~1\times 10^9 $ & 50,~~200\\
$M_{31,P2}$ & 28 &    2.297~81~( 13) & $ 1\times 10^8,~~1\times 10^9 $ & 50,~~180\\
$M_{32,P2}$ & 56 &   -2.675~78~( 35) & $ 1\times 10^8,~~1\times 10^9 $ & 50,~~180\\
$M_{33,P2}$ & 28 &   -0.960~21~(  6) & $ 1\times 10^8,~~1\times 10^9 $ & 50,~~180\\
$M_{34,P2}$ & 56 &   -0.967~04~( 34) & $ 1\times 10^8,~~1\times 10^9 $ & 50,~~180\\
$M_{35,P2}$ & 56 &   -0.796~99~( 36) & $ 1\times 10^8,~~1\times 10^9 $ & 50,~~180\\
$M_{36,P2}$ & 56 &    1.171~39~( 41) & $ 1\times 10^8,~~1\times 10^9 $ & 50,~~180\\
$M_{37,P2}$ & 28 &    0.709~94~( 13) & $ 1\times 10^8,~~1\times 10^9 $ & 50,~~180\\
$M_{38,P2}$ & 28 &    0.247~72~( 29) & $ 1\times 10^8,~~1\times 10^9 $ & 50,~~200\\
$M_{39,P2}$ & 56 &   -0.830~00~( 30) & $ 1\times 10^8,~~1\times 10^9 $ & 50,~~180\\
$M_{40,P2}$ & 56 &   -0.499~07~( 47) & $ 1\times 10^8,~~1\times 10^9 $ & 50,~~200\\
$M_{41,P2}$ & 28 &   -1.083~44~( 71) & $ 1\times 10^8,~~1\times 10^9 $ & 50,~~200\\
$M_{42,P2}$ & 28 &    0.576~12~( 76) & $ 1\times 10^8,~~1\times 10^9 $ & 50,~~200\\
$M_{43,P2}$ & 28 &   -1.074~51~( 41) & $ 1\times 10^8,~~1\times 10^9 $ & 50,~~200\\
$M_{44,P2}$ & 56 &    1.919~47~( 60) & $ 1\times 10^8,~~1\times 10^9 $ & 50,~~200\\
$M_{45,P2}$ & 28 &    0.011~51~( 37) & $ 1\times 10^8,~~1\times 10^9 $ & 50,~~200\\
$M_{46,P2}$ & 28 &   -0.588~88~( 73) & $ 1\times 10^8,~~1\times 10^9 $ & 50,~~200\\
$M_{47,P2}$ & 28 &   -0.102~58~( 65) & $ 1\times 10^8,~~1\times 10^9 $ & 50,~~200\\

  \end{tabular}
  \end{ruledtabular}
\end{table*}

\endgroup

%% file: res-tableEe.tex
\begin{table}
\caption{
Residual renormalization constants needed for the calculation of
$a_e^{(10)}[ {\rm Set IV}^{(ee)}]$. 
Notations are those of Eq.~(\ref{a8_setIV}).
 \label{Table:residual_const}
}
\begin{ruledtabular}

\begin{tabular}{l@{\hskip-3em}dl@{\hskip-3em}d}
    $ \Delta M_{6,P2}$   &           1.014~060~(30) &
    $ \Delta M_6$        &           0.425~820~(14)   \\
    $ \Delta M_{4,P2}$   &          -0.106~707\cdots   &
    $ \Delta M_{4}$      &           0.030~833\cdots   \\
    $ M_{2,P2}$   &           0.015~687\cdots    &
    $ M_{2}$      &           0.5                \\
    $ \Delta \LB_{6,P2}$  &           0.351~54  ~  (93)   &
    $ \Delta \LB_{6}$     &           0.100~86  ~  (77)    \\
    $ \Delta \LB_{4,P2}$ &          -0.114~228 ~  (17)     &
    $ \Delta \LB_{4}$    &           0.027~930~   (27)     \\
    $ \Delta \delta m_{4,P2}$ &      0.679~769 ~  (15)          &
    $ \Delta \delta m_{4}$    &      1.906~340  ~ (21)          \\
    $ \Delta \LB_{2,P2}$ &           0.063~399\cdots        &
    $ \Delta \LB_{2}$    &           0.75                   \\
    $ \Delta L_{2^*,P2}$ &          -0.023~531\cdots         &
    $ \Delta L_{2^*}$    &          -0.75                   \\
    $ \Delta B_{2^*,P2}$ &           0.047~062\cdots         &
    $ \Delta B_{2^*}$    &           1.5                    \\
\end{tabular}
\end{ruledtabular}
\end{table}


%% file: tableEm.tex
\begingroup
\renewcommand{\baselinestretch}{1.0}

\begin{table*}
  \caption{%
Contributions of diagrams $M_{01,P2}$,\dots, $M_{24,P2}$ of Set~IV to $a_e$ 
for $(l_1l_2) = (em)$. 
The multiplicity $n_F$ is the number of vertex diagrams 
represented by the integral and 
is incorporated in the numerical value. 
The integral $M_{12,P2}$ is evaluated with quadruple 
precision.
All other integrals are evaluated with double precision.
  \label{table:setIV_em_1}
}

\begin{ruledtabular}
    \begin{tabular}{lcdcc}
\multicolumn{1}{c}{Integral} & 
\multicolumn{1}{c}{$n_F$} &
\multicolumn{1}{c}{\hspace*{4em}\parbox[t]{10em}{Value (Error) \\[-1ex] including $n_F$}} &
\multicolumn{1}{c}{\parbox[t]{10em}{Sampling per \\[-1ex]  iteration}} &
\multicolumn{1}{c}{\parbox[t]{4em}{No. of \\[-1ex] iterations}} \\[3ex]
\hline

$M_{01,P2}^{(em)}$ & 28 &    0.000~759~(  3) & $ 1\times 10^7,~1\times 10^8 $ &100,~20\\
$M_{02,P2}^{(em)}$ & 56 &    0.000~205~(  8) & $ 1\times 10^7,~1\times 10^8 $ &100,~20\\
$M_{03,P2}^{(em)}$ & 28 &    0.001~757~(  6) & $ 1\times 10^7,~1\times 10^8 $ &100,~20\\
$M_{04,P2}^{(em)}$ & 56 &    0.001~170~( 15) & $ 1\times 10^7,~1\times 10^8 $ &100,~20\\
$M_{05,P2}^{(em)}$ & 56 &    0.000~197~(  7) & $ 1\times 10^7,~1\times 10^8 $ &100,~20\\
$M_{06,P2}^{(em)}$ & 56 &   -0.001~845~( 12) & $ 1\times 10^7,~1\times 10^8 $ &100,~20\\
$M_{07,P2}^{(em)}$ & 56 &   -0.001~866~(  5) & $ 1\times 10^7,~1\times 10^8 $ &100,~20\\
$M_{08,P2}^{(em)}$ & 56 &    0.000~922~( 11) & $ 1\times 10^7,~1\times 10^8 $ &100,~20\\
$M_{09,P2}^{(em)}$ & 56 &   -0.001~077~( 16) & $ 1\times 10^7,~1\times 10^8 $ &100,~20\\
$M_{10,P2}^{(em)}$ & 56 &   -0.001~316~(  9) & $ 1\times 10^7,~1\times 10^8 $ &100,~20\\
$M_{11,P2}^{(em)}$ & 28 &    0.000~310~(  3) & $ 1\times 10^7,~1\times 10^8 $ &100,~20\\
$M_{12,P2}^{(em)}$ & 28 &    0.000~822~(  1) & $ 1\times 10^7               $ &20     \\
$M_{13,P2}^{(em)}$ & 56 &   -0.001~434~(  8) & $ 1\times 10^7,~1\times 10^8 $ &100,~20\\
$M_{14,P2}^{(em)}$ & 56 &    0.000~301~( 10) & $ 1\times 10^7,~1\times 10^8 $ &100,~20\\
$M_{15,P2}^{(em)}$ & 56 &    0.000~141~(  5) & $ 1\times 10^7,~1\times 10^8 $ &100,~20\\
$M_{16,P2}^{(em)}$ & 56 &    0.000~920~(  8) & $ 1\times 10^7,~1\times 10^8 $ &100,~20\\
$M_{17,P2}^{(em)}$ & 56 &    0.002~263~( 14) & $ 1\times 10^7,~1\times 10^8 $ &100,~20\\
$M_{18,P2}^{(em)}$ & 56 &    0.000~285~(  4) & $ 1\times 10^7,~1\times 10^8 $ &100,~20\\
$M_{19,P2}^{(em)}$ & 28 &   -0.001~671~(  1) & $ 1\times 10^7,~1\times 10^8 $ &100,~20\\
$M_{20,P2}^{(em)}$ & 56 &    0.000~738~( 12) & $ 1\times 10^7,~1\times 10^8 $ &100,~20\\
$M_{21,P2}^{(em)}$ & 28 &    0.000~210~(  1) & $ 1\times 10^7,~1\times 10^8 $ &100,~20\\
$M_{22,P2}^{(em)}$ & 56 &    0.000~880~(  8) & $ 1\times 10^7,~1\times 10^8 $ &100,~20\\
$M_{23,P2}^{(em)}$ & 56 &    0.000~105~( 27) & $ 1\times 10^7,~1\times 10^8 $ &100,~20\\
$M_{24,P2}^{(em)}$ & 56 &    0.000~371~( 11) & $ 1\times 10^7,~1\times 10^8 $ &100,~20\\

  \end{tabular}
  \end{ruledtabular}
\end{table*}
\endgroup

\begingroup
\renewcommand{\baselinestretch}{1.0}

\begin{table*}
  \caption{%
Contributions of diagrams $M_{24,P2}$,\dots, $M_{47,P2}$ of Set~IV to $a_e$ 
for $(l_1l_2) = (em)$. 
The multiplicity $n_F$ is the number of vertex diagrams 
represented by the integral and 
is incorporated in the numerical value. 
All integrals are evaluated with double precision.
  \label{table:setIV_em_2}
}

\begin{ruledtabular}
    \begin{tabular}{lcdcc}
\multicolumn{1}{c}{Integral} & 
\multicolumn{1}{c}{$n_F$} &
\multicolumn{1}{c}{\hspace*{4em}\parbox[t]{10em}{Value (Error) \\[-1ex] including $n_F$}} &
\multicolumn{1}{c}{\parbox[t]{10em}{Sampling per \\[-1ex]  iteration}} &
\multicolumn{1}{c}{\parbox[t]{4em}{No. of \\[-1ex] iterations}} \\[3ex]
\hline

$M_{25,P2}^{(em)}$ & 28 &   -0.001~089~(  4) & $ 1\times 10^7,~1\times 10^8 $ &100,~20\\
$M_{26,P2}^{(em)}$ & 28 &    0.000~247~(  3) & $ 1\times 10^7,~1\times 10^8 $ &100,~20\\
$M_{27,P2}^{(em)}$ & 56 &    0.001~937~( 15) & $ 1\times 10^7,~1\times 10^8 $ &100,~20\\
$M_{28,P2}^{(em)}$ & 56 &    0.000~000~(  8) & $ 1\times 10^7,~1\times 10^8 $ &100,~20\\
$M_{29,P2}^{(em)}$ & 28 &    0.000~786~( 10) & $ 1\times 10^7,~1\times 10^8 $ &100,~20\\
$M_{30,P2}^{(em)}$ & 28 &    0.000~014~(  4) & $ 1\times 10^7,~1\times 10^8 $ &100,~20\\
$M_{31,P2}^{(em)}$ & 28 &    0.001~088~(  3) & $ 1\times 10^7,~1\times 10^8 $ &100,~20\\
$M_{32,P2}^{(em)}$ & 56 &   -0.002~282~(  9) & $ 1\times 10^7,~1\times 10^8 $ &100,~20\\
$M_{33,P2}^{(em)}$ & 28 &    0.000~774~(  1) & $ 1\times 10^7,~1\times 10^8 $ &100,~20\\
$M_{34,P2}^{(em)}$ & 56 &   -0.001~448~(  7) & $ 1\times 10^7,~1\times 10^8 $ &100,~20\\
$M_{35,P2}^{(em)}$ & 56 &    0.000~296~(  8) & $ 1\times 10^7,~1\times 10^8 $ &100,~20\\
$M_{36,P2}^{(em)}$ & 56 &    0.001~241~(  8) & $ 1\times 10^7,~1\times 10^8 $ &100,~20\\
$M_{37,P2}^{(em)}$ & 28 &    0.000~421~(  3) & $ 1\times 10^7,~1\times 10^8 $ &100,~20\\
$M_{38,P2}^{(em)}$ & 28 &    0.000~563~(  3) & $ 1\times 10^7,~1\times 10^8 $ &100,~20\\
$M_{39,P2}^{(em)}$ & 56 &    0.000~892~(  5) & $ 1\times 10^7,~1\times 10^8 $ &100,~20\\
$M_{40,P2}^{(em)}$ & 56 &    0.000~409~(  3) & $ 1\times 10^7,~1\times 10^8 $ &100,~20\\
$M_{41,P2}^{(em)}$ & 28 &    0.000~775~(  7) & $ 1\times 10^7,~1\times 10^8 $ &100,~20\\
$M_{42,P2}^{(em)}$ & 28 &    0.000~079~(  4) & $ 1\times 10^7,~1\times 10^8 $ &100,~20\\
$M_{43,P2}^{(em)}$ & 28 &    0.000~246~( 11) & $ 1\times 10^7,~1\times 10^8 $ &100,~20\\
$M_{44,P2}^{(em)}$ & 56 &    0.000~546~( 10) & $ 1\times 10^7,~1\times 10^8 $ &100,~20\\
$M_{45,P2}^{(em)}$ & 28 &    0.000~117~(  2) & $ 1\times 10^7,~1\times 10^8 $ &100,~20\\
$M_{46,P2}^{(em)}$ & 28 &    0.000~674~(  8) & $ 1\times 10^7,~1\times 10^8 $ &100,~20\\
$M_{47,P2}^{(em)}$ & 28 &    0.000~203~(  4) & $ 1\times 10^7,~1\times 10^8 $ &100,~20\\

  \end{tabular}
  \end{ruledtabular}
\end{table*}

\endgroup

%% file: res-tableMe.tex
\begin{table}
\caption{
Residual renormalization constants needed for the calculation of
the mass-dependent contributions from Set~IV diagrams.
Notations are those of Eq.~(\ref{a8_setIV}).
 \label{Table:residual_const_mass_dep}
}
\begin{ruledtabular}

\begin{tabular}{l@{\hskip-3em}dl@{\hskip-3em}d}
    $ \Delta M_{6,P2}^{(em)}$        &   0.000~721~65~(94)       &
    $ \Delta M_{4,P2}^{(em)}$        &  -0.000~018~910~(26)     \\
    $ M_{2,P2}^{(em)}$        &   0.000~000~519~762~(21)  &
                                     &                     \\
    $ \Delta \LB_{6,P2}^{(em)}$      &   0.000~705~(12)       & 
    $ \Delta \LB_{4,P2}^{(em)}$      &  -0.000~079~83~(10)     \\ 
    $ \Delta \delta m_{4,P2}^{(em)}$ &   0.000~255~64~(5)     & 
    $ \Delta \LB_{2,P2}^{(em)}$      &   0.000~009~405~25~(83)    \\   
    $ \Delta  L_{2^*,P2}^{(em)}$     &  -0.000~000~779~612~(11)    & 
    $ \Delta  B_{2^*,P2}^{(em)}$     &   0.000~001~559~224~(19)   \\ 
                &              &     &                         \\
    $ \Delta M_{6,P2}^{(me)}$        &   5.374~0~(45)      &
    $ \Delta M_{4,P2}^{(me)}$        &  -0.628~832\cdots   \\
    $ M_{2,P2}^{(me)}$        &   1.094~258\cdots   &
                                     &                     \\
    $ \Delta \LB_{6,P2}^{(me)}$      &   1.476~3~(33)       & 
    $ \Delta \LB_{4,P2}^{(me)}$      &  -0.308~75~(32)     \\ 
    $ \Delta \delta m_{4,P2}^{(me)}$ &  11.151~39~(32)     & 
    $ \Delta \LB_{2,P2}^{(me)}$      &   1.885~733~(16)    \\   
    $ \Delta  L_{2^*,P2}^{(me)}$     &  -1.641~436~(54)    & 
    $ \Delta  B_{2^*,P2}^{(me)}$     &   3.282~872~(107)   \\ 
                &              &     &                         \\
    $ \Delta M_{6,P2}^{(mt)}$        &   0.038~01~(14)      &
    $ \Delta M_{4,P2}^{(mt)}$        &  -0.001~641~9~(18)   \\
    $ M_{2,P2}^{(mt)}$        &   0.000~078~067~4~(31)   &
                                     &                     \\
    $ \Delta \LB_{6,P2}^{(mt)}$      &   0.023~97~(29)       & 
    $ \Delta \LB_{4,P2}^{(mt)}$      &  -0.004~155~7~(45)    \\ 
    $ \Delta \delta m_{4,P2}^{(mt)}$ &   0.015~483~(25)       & 
    $ \Delta \LB_{2,P2}^{(mt)}$      &   0.000~831~107~(75)   \\   
    $ \Delta  L_{2^*,P2}^{(mt)}$     &  -0.000~117~097~0~(15)  & 
    $ \Delta  B_{2^*,P2}^{(mt)}$     &   0.000~234~(1)         \\ 
\end{tabular}
\end{ruledtabular}
\end{table}

%% file: tableMe.tex
\begingroup
\renewcommand{\baselinestretch}{1.0}

\begin{table*}
  \caption{%
Contributions of diagrams $M_{01,P2}$,\dots, $M_{24,P2}$ of Set~IV to $a_\mu$ 
for $(l_1l_2) = (me)$. 
The multiplicity $n_F$ is the number of vertex diagrams 
represented by the integral and 
is incorporated in the numerical value. 
The integrals $M_{12,P2}$, $M_{16,P2}$, and $M_{18,P2}$ are evaluated with quadruple precision.
All other integrals are evaluated with double precision.
  \label{table:setIV_me_1}
}

\begin{ruledtabular}
    \begin{tabular}{lcdcc}
\multicolumn{1}{c}{Integral} & 
\multicolumn{1}{c}{$n_F$} &
\multicolumn{1}{c}{\hspace*{4em}\parbox[t]{10em}{Value (Error) \\[-1ex] including $n_F$}} &
\multicolumn{1}{c}{\parbox[t]{10em}{Sampling per \\[-1ex]  iteration}} &
\multicolumn{1}{c}{\parbox[t]{4em}{No. of \\[-1ex] iterations}} \\[3ex]
\hline

$M_{01,P2}^{(me)} $ &28 &  -0.369~5~(130) & $ 1\times 10^8,~~1\times 10^9 $ & 80,~~ 48\\
$M_{02,P2}^{(me)} $ &56 &  -8.223~2~(308) & $ 1\times 10^8,~~1\times 10^9 $ & 80,~~140\\
$M_{03,P2}^{(me)} $ &28 &  -3.986~6~(230) & $ 1\times 10^8,~~1\times 10^9 $ & 80,~~ 80\\
$M_{04,P2}^{(me)} $ &56 &  46.429~2~(587) & $ 1\times 10^8,~~1\times 10^9 $ & 80,~~156\\
$M_{05,P2}^{(me)} $ &56 &  19.803~9~(105) & $ 1\times 10^8,~~1\times 10^9 $ & 80,~~ 48\\
$M_{06,P2}^{(me)} $ &56 & -11.614~8~(211) & $ 1\times 10^8,~~1\times 10^9 $ & 80,~~ 80\\
$M_{07,P2}^{(me)} $ &56 &  -0.558~3~(129) & $ 1\times 10^8,~~1\times 10^9 $ & 80,~~ 72\\
$M_{08,P2}^{(me)} $ &56 & -48.877~7~(420) & $ 1\times 10^8,~~1\times 10^9 $ & 80,~~130\\
$M_{09,P2}^{(me)} $ &56 &   4.817~2~(310) & $ 1\times 10^8,~~1\times 10^9 $ & 80,~~140\\
$M_{10,P2}^{(me)} $ &56 &  18.291~5~(459) & $ 1\times 10^8,~~1\times 10^9 $ & 80,~~156\\
$M_{11,P2}^{(me)} $ &28 &  21.337~7~(299) & $ 1\times 10^8,~~1\times 10^9 $ & 80,~~100\\
$M_{12,P2}^{(me)} $ &28 & -56.967~6~(174) & $ 1\times 10^8,~~1\times 10^9 $ & 50,~~ 20\\
$M_{13,P2}^{(me)} $ &56 & -61.803~8~(142) & $ 1\times 10^8,~~1\times 10^9 $ & 80,~~ 48\\
$M_{14,P2}^{(me)} $ &56 &  21.147~2~(238) & $ 1\times 10^8,~~1\times 10^9 $ & 80,~~ 80\\
$M_{15,P2}^{(me)} $ &56 &   7.639~9~(138) & $ 1\times 10^8,~~1\times 10^9 $ & 80,~~ 72\\
$M_{16,P2}^{(me)} $ &56 &  62.954~8~(414) & $ 1\times 10^8,~~1\times 10^9 $ & 50,~~ 35\\
$M_{17,P2}^{(me)} $ &56 &  62.823~6~(412) & $ 1\times 10^8,~~1\times 10^9 $ & 80,~~156\\
$M_{18,P2}^{(me)} $ &56 & -44.191~1~(207) & $ 1\times 10^8,~~1\times 10^9 $ & 50,~~ 30\\
$M_{19,P2}^{(me)} $ &28 & -12.057~1~( 14) & $ 1\times 10^8,~~1\times 10^9 $ & 80,~~ 40\\
$M_{20,P2}^{(me)} $ &56 &   9.281~7~( 84) & $ 1\times 10^8,~~1\times 10^9 $ & 80,~~ 48\\
$M_{21,P2}^{(me)} $ &28 &   4.359~0~( 12) & $ 1\times 10^8,~~1\times 10^9 $ & 80,~~ 40\\
$M_{22,P2}^{(me)} $ &56 &  -2.934~2~(105) & $ 1\times 10^8,~~1\times 10^9 $ & 80,~~ 48\\
$M_{23,P2}^{(me)} $ &56 & -44.431~4~(185) & $ 1\times 10^8,~~1\times 10^9 $ & 80,~~ 72\\
$M_{24,P2}^{(me)} $ &56 &  19.396~5~(175) & $ 1\times 10^8,~~1\times 10^9 $ & 80,~~ 72\\
  \end{tabular}
  \end{ruledtabular}
\end{table*}

\endgroup

\begingroup
\renewcommand{\baselinestretch}{1.0}

\begin{table*}
  \caption{%
Contributions of diagrams $M_{25,P2}$,\dots, $M_{47,P2}$ of Set~IV to $a_\mu$ 
for $(l_1l_2) = (me)$. 
The multiplicity $n_F$ is the number of vertex diagrams 
represented by the integral and 
is incorporated in the numerical value. 
All integrals are evaluated with double precision.
  \label{table:setIV_me_2}
}

\begin{ruledtabular}
    \begin{tabular}{lcdcc}
\multicolumn{1}{c}{Integral} & 
\multicolumn{1}{c}{$n_F$} &
\multicolumn{1}{c}{\hspace*{4em}\parbox[t]{10em}{Value (Error) \\[-1ex] including $n_F$}} &
\multicolumn{1}{c}{\parbox[t]{10em}{Sampling per \\[-1ex]  iteration}} &
\multicolumn{1}{c}{\parbox[t]{4em}{No. of \\[-1ex] iterations}} \\[3ex]
\hline

$M_{25,P2}^{(me)} $ &28 &  -1.148~1~( 73) & $ 1\times 10^8,~~1\times 10^9 $ & 80,~~ 40\\
$M_{26,P2}^{(me)} $ &28 & -13.572~5~(177) & $ 1\times 10^8,~~1\times 10^9 $ & 80,~~ 80\\
$M_{27,P2}^{(me)} $ &56 &  11.510~8~(246) & $ 1\times 10^8,~~1\times 10^9 $ & 80,~~ 80\\
$M_{28,P2}^{(me)} $ &56 &  36.510~0~(394) & $ 1\times 10^8,~~1\times 10^9 $ & 80,~~156\\
$M_{29,P2}^{(me)} $ &28 &  36.421~2~(298) & $ 1\times 10^8,~~1\times 10^9 $ & 80,~~ 88\\
$M_{30,P2}^{(me)} $ &28 & -43.675~1~(402) & $ 1\times 10^8,~~1\times 10^9 $ & 80,~~156\\
$M_{31,P2}^{(me)} $ &28 &  35.547~0~( 22) & $ 1\times 10^8,~~1\times 10^9 $ & 80,~~ 40\\
$M_{32,P2}^{(me)} $ &56 & -30.768~6~( 60) & $ 1\times 10^8,~~1\times 10^9 $ & 80,~~ 40\\
$M_{33,P2}^{(me)} $ &28 & -14.328~3~( 11) & $ 1\times 10^8,~~1\times 10^9 $ & 80,~~ 40\\
$M_{34,P2}^{(me)} $ &56 &   8.112~3~( 66) & $ 1\times 10^8,~~1\times 10^9 $ & 80,~~ 40\\
$M_{35,P2}^{(me)} $ &56 &  -7.263~8~( 65) & $ 1\times 10^8,~~1\times 10^9 $ & 80,~~ 40\\
$M_{36,P2}^{(me)} $ &56 &   3.414~7~( 87) & $ 1\times 10^8,~~1\times 10^9 $ & 80,~~ 48\\
$M_{37,P2}^{(me)} $ &28 &   7.820~6~( 24) & $ 1\times 10^8,~~1\times 10^9 $ & 80,~~ 40\\
$M_{38,P2}^{(me)} $ &28 & -16.252~5~( 81) & $ 1\times 10^8,~~1\times 10^9 $ & 80,~~ 48\\
$M_{39,P2}^{(me)} $ &56 &  -7.644~5~( 78) & $ 1\times 10^8,~~1\times 10^9 $ & 80,~~ 40\\
$M_{40,P2}^{(me)} $ &56 &   2.819~0~(158) & $ 1\times 10^8,~~1\times 10^9 $ & 80,~~ 72\\
$M_{41,P2}^{(me)} $ &28 & -25.777~5~(194) & $ 1\times 10^8,~~1\times 10^9 $ & 80,~~ 72\\
$M_{42,P2}^{(me)} $ &28 &  26.504~0~(291) & $ 1\times 10^8,~~1\times 10^9 $ & 80,~~ 88\\
$M_{43,P2}^{(me)} $ &28 & -29.852~0~(100) & $ 1\times 10^8,~~1\times 10^9 $ & 80,~~ 48\\
$M_{44,P2}^{(me)} $ &56 &  37.460~8~(173) & $ 1\times 10^8,~~1\times 10^9 $ & 80,~~ 72\\
$M_{45,P2}^{(me)} $ &28 &   9.735~1~(173) & $ 1\times 10^8,~~1\times 10^9 $ & 80,~~ 72\\
$M_{46,P2}^{(me)} $ &28 &   8.399~9~(213) & $ 1\times 10^8,~~1\times 10^9 $ & 80,~~ 80\\
$M_{47,P2}^{(me)} $ &28 & -22.410~1~(284) & $ 1\times 10^8,~~1\times 10^9 $ & 80,~~104\\

  \end{tabular}
  \end{ruledtabular}
\end{table*}

\endgroup

%% file: tableMt.tex
\begingroup
\renewcommand{\baselinestretch}{1.0}

\begin{table*}
  \caption{%
Contributions of diagrams $M_{01,P2}$,\dots, $M_{24,P2}$ of Set~IV to $a_\mu$ 
for $(l_1l_2) = (mt)$. 
The multiplicity $n_F$ is the number of vertex diagrams 
represented by the integral and 
is incorporated in the numerical value. 
The integral $M_{12,P2}$ is evaluated with quadruple
precision.
All other integrals are evaluated with double precision.
  \label{table:setIV_mt_1}
}

\begin{ruledtabular}
    \begin{tabular}{lcdcc}
\multicolumn{1}{c}{Integral} & 
\multicolumn{1}{c}{$n_F$} &
\multicolumn{1}{c}{\hspace*{4em}\parbox[t]{10em}{Value (Error) \\[-1ex] including $n_F$}} &
\multicolumn{1}{c}{\parbox[t]{10em}{Sampling per \\[-1ex]  iteration}} &
\multicolumn{1}{c}{\parbox[t]{4em}{No. of \\[-1ex] iterations}} \\[3ex]
\hline
$M_{01,P2}^{(mt)}$ & 28 &   -0.026~89~( 15) & $ 1\times 10^7 $ &100\\
$M_{02,P2}^{(mt)}$ & 56 &   -0.002~06~( 37) & $ 1\times 10^7 $ &100\\
$M_{03,P2}^{(mt)}$ & 28 &    0.051~83~( 23) & $ 1\times 10^7 $ &100\\
$M_{04,P2}^{(mt)}$ & 56 &    0.049~99~( 55) & $ 1\times 10^7 $ &100\\
$M_{05,P2}^{(mt)}$ & 56 &   -0.011~20~( 29) & $ 1\times 10^7 $ &100\\
$M_{06,P2}^{(mt)}$ & 56 &   -0.060~08~( 52) & $ 1\times 10^7 $ &100\\
$M_{07,P2}^{(mt)}$ & 56 &   -0.065~38~( 24) & $ 1\times 10^7 $ &100\\
$M_{08,P2}^{(mt)}$ & 56 &   -0.038~51~( 55) & $ 1\times 10^7 $ &100\\
$M_{09,P2}^{(mt)}$ & 56 &   -0.026~52~( 62) & $ 1\times 10^7 $ &100\\
$M_{10,P2}^{(mt)}$ & 56 &   -0.050~11~( 44) & $ 1\times 10^7 $ &100\\
$M_{11,P2}^{(mt)}$ & 28 &    0.016~61~( 16) & $ 1\times 10^7 $ &100\\
$M_{12,P2}^{(mt)}$ & 28 &   -0.037~85~(  2) & $ 1\times 10^7 $ & 50\\
$M_{13,P2}^{(mt)}$ & 56 &   -0.060~34~( 30) & $ 1\times 10^7 $ &100\\
$M_{14,P2}^{(mt)}$ & 56 &    0.000~86~( 43) & $ 1\times 10^7 $ &100\\
$M_{15,P2}^{(mt)}$ & 56 &    0.009~38~( 23) & $ 1\times 10^7 $ &100\\
$M_{16,P2}^{(mt)}$ & 56 &    0.040~26~( 49) & $ 1\times 10^7 $ &100\\
$M_{17,P2}^{(mt)}$ & 56 &    0.101~13~( 62) & $ 1\times 10^7 $ &100\\
$M_{18,P2}^{(mt)}$ & 56 &    0.011~93~( 28) & $ 1\times 10^7 $ &100\\
$M_{19,P2}^{(mt)}$ & 28 &   -0.063~10~(  5) & $ 1\times 10^7 $ &100\\
$M_{20,P2}^{(mt)}$ & 56 &    0.029~21~( 37) & $ 1\times 10^7 $ &100\\
$M_{21,P2}^{(mt)}$ & 28 &    0.007~11~(  4) & $ 1\times 10^7 $ &100\\
$M_{22,P2}^{(mt)}$ & 56 &   -0.034~71~( 33) & $ 1\times 10^7 $ &100\\
$M_{23,P2}^{(mt)}$ & 56 &   -0.008~27~( 68) & $ 1\times 10^7 $ &100\\
$M_{24,P2}^{(mt)}$ & 56 &    0.022~23~( 39) & $ 1\times 10^7 $ &100\\
  \end{tabular}
  \end{ruledtabular}
\end{table*}

\endgroup

\begingroup
\renewcommand{\baselinestretch}{1.0}

\begin{table*}
  \caption{%
Contributions of diagrams $M_{25,P2}$,\dots, $M_{47,P2}$ of Set~IV to $a_\mu$ 
for $(l_1l_2) = (mt)$. 
The multiplicity $n_F$ is the number of vertex diagrams 
represented by the integral and 
is incorporated in the numerical value. 
All integrals are evaluated with double precision.
  \label{table:setIV_mt_2}
}

\begin{ruledtabular}
    \begin{tabular}{lcdcc}
\multicolumn{1}{c}{Integral} & 
\multicolumn{1}{c}{$n_F$} &
\multicolumn{1}{c}{\hspace*{4em}\parbox[t]{10em}{Value (Error) \\[-1ex] including $n_F$}} &
\multicolumn{1}{c}{\parbox[t]{10em}{Sampling per \\[-1ex]  iteration}} &
\multicolumn{1}{c}{\parbox[t]{4em}{No. of \\[-1ex] iterations}} \\[3ex]
\hline
$M_{25,P2}^{(mt)}$ & 28 &   -0.039~19~( 15) & $ 1\times 10^7 $ &100\\
$M_{26,P2}^{(mt)}$ & 28 &   -0.010~15~( 17) & $ 1\times 10^7 $ &100\\
$M_{27,P2}^{(mt)}$ & 56 &    0.060~33~( 53) & $ 1\times 10^7 $ &100\\
$M_{28,P2}^{(mt)}$ & 56 &    0.006~03~( 36) & $ 1\times 10^7 $ &100\\
$M_{29,P2}^{(mt)}$ & 28 &    0.037~86~( 34) & $ 1\times 10^7 $ &100\\
$M_{30,P2}^{(mt)}$ & 28 &   -0.005~19~( 24) & $ 1\times 10^7 $ &100\\
$M_{31,P2}^{(mt)}$ & 28 &    0.056~43~( 12) & $ 1\times 10^7 $ &100\\
$M_{32,P2}^{(mt)}$ & 56 &   -0.096~26~( 30) & $ 1\times 10^7 $ &100\\
$M_{33,P2}^{(mt)}$ & 28 &   -0.030~72~(  5) & $ 1\times 10^7 $ &100\\
$M_{34,P2}^{(mt)}$ & 56 &   -0.060~57~( 25) & $ 1\times 10^7 $ &100\\
$M_{35,P2}^{(mt)}$ & 56 &   -0.025~98~( 31) & $ 1\times 10^7 $ &100\\
$M_{36,P2}^{(mt)}$ & 56 &    0.050~76~( 32) & $ 1\times 10^7 $ &100\\
$M_{37,P2}^{(mt)}$ & 28 &    0.019~70~( 10) & $ 1\times 10^7 $ &100\\
$M_{38,P2}^{(mt)}$ & 28 &    0.021~52~( 14) & $ 1\times 10^7 $ &100\\
$M_{39,P2}^{(mt)}$ & 56 &   -0.034~44~( 21) & $ 1\times 10^7 $ &100\\
$M_{40,P2}^{(mt)}$ & 56 &   -0.018~44~( 21) & $ 1\times 10^7 $ &100\\
$M_{41,P2}^{(mt)}$ & 28 &   -0.033~08~( 28) & $ 1\times 10^7 $ &100\\
$M_{42,P2}^{(mt)}$ & 28 &    0.006~74~( 25) & $ 1\times 10^7 $ &100\\
$M_{43,P2}^{(mt)}$ & 28 &   -0.016~64~( 36) & $ 1\times 10^7 $ &100\\
$M_{44,P2}^{(mt)}$ & 56 &    0.039~39~( 35) & $ 1\times 10^7 $ &100\\
$M_{45,P2}^{(mt)}$ & 28 &    0.002~23~( 14) & $ 1\times 10^7 $ &100\\
$M_{46,P2}^{(mt)}$ & 28 &   -0.028~76~( 34) & $ 1\times 10^7 $ &100\\
$M_{47,P2}^{(mt)}$ & 28 &   -0.004~02~( 21) & $ 1\times 10^7 $ &100\\
  \end{tabular}
  \end{ruledtabular}
\end{table*}

\endgroup

%% file: appA.tex
The purpose of this Appendix is to obtain the sum of residual
renormalization terms of the Set IV.
Since diagrams of Set IV have exact correspondence with
the diagrams of Group V of the {\it eighth-order} $g\!-\!2$,
however, it is simpler to consider the residual renormalization of the diagrams of Group V,
from which the residual renormalization of the Set IV can be readily derived.

In our approach integrals of individual diagrams must be made convergent
before they are integrated numerically.
This is achieved by constructing terms which subtract UV-divergent parts
by {\it K}-operation and IR-divergent parts by {\it I/R}-subtractions.
Since this scheme is different from the standard on-shell renormalization,
it is necessary to make an adjustment, called residual renormalization.
Residual renormalization terms of individual diagrams must then be
summed up over all diagrams involved.

As the order of perturbation increases the total number of terms 
contributing to the residual renormalization increases rapidly
so that it will become harder and harder to manage.
Fortunately the sum of all residual terms can be expressed concisely
in terms of magnetic moments and finite parts of renormalization
constants of lower orders \cite{Aoyama:2007mn}, 
and the sum has a structure closely related to that of
the standard on-shell renormalization.
This enables us to confirm the validity of
the sum of residual renormalization terms
starting from the expression of the standard renormalization.

To see this relation clearly it is useful
to treat UV-divergence and IR-divergence separately.
We present the logic of our approach for the fourth-, sixth-, 
and eighth-order cases, in that order.
We deal here only with 
Ward-Takahashi(WT)-summed diagrams of $q$-type, namely diagrams without closed lepton loops.
Thus $M_{2n}$ and  $a_{2n}$, $n=1,2,\cdots$, refer to 
unrenormalized and renormalized 
amplitudes of such diagrams, respectively.

Our discussion here follows the scheme incorporated in the
automatic code generator {\sc gencode}{\it N},
which is applicable to any value of the order {\it N}.


\subsection{fourth-order case}

The standard renormalization of the fourth-order magnetic moment $a_4$ 
can be expressed in the form
\begin{equation}
a_4 = M_4 - 2 L_2 M_2 - B_2 M_2 - {\delta m}_2 M_{2^*},              
\label{a4_standard}
\end{equation}
where $M_2$ is the second-order magnetic moment,
$M_{2^*}$ is obtained from $M_2$ by inserting a two-point vertex
in the lepton line of $M_2$, and
$M_4$ is the sum of unrenormalized WT-summed amplitudes $M_{4a}$ and $M_{4b}$:
\begin{equation}
M_4\equiv M_{4a}+M_{4b},
\label{defM4}
\end{equation}
where $4a$ and $4b$ refer to the fourth-order diagrams in which
two virtual photons are crossed and uncrossed, respectively.
The coefficients of renormalization constants $L_2$ and $B_2$ in 
Eq.~(\ref{a4_standard}) reflect the fact that $M_{4a}$ is obtained
by inserting a second-order vertex diagram in two vertices of $M_2$ and
$M_{4b}$ is obtained by inserting a
second-order self-energy diagram in the electron line of $M_2$.


\subsubsection{Separation of UV divergences by the {\it K}-operation}

$M_4$ has no {\it overall} UV divergence.
However, it has UV divergences coming from subdiagrams.
Applying {\it K}-operation on these divergences we obtain
\begin{equation}
     M_4 = B^{\rm UV}_2 M_2 + {\delta m}^{\rm UV}_2 M_{2^*} + 2 L^{\rm UV}_2 M_2 + M_4^{\rm R} , 
\label{M4}
\end{equation}
where the superscript ${\rm R}$ in $M_4^{\rm R}$ means that all 
subdiagram UV divergences
are removed from $M_4$.
$L^{\rm UV}_2$ and $B^{\rm UV}_2$ are the UV-divergent parts 
separated out from $L_2$ and $B_2$ by the {\it K}-operation
and ${L}_2^{\rm R}$ and ${B}_2^{\rm R}$ are UV-finite 
(but IR-divergent) remainders: 
\begin{eqnarray}
     L_2 &=& L^{\rm UV}_2 + {L}_2^{\rm R} ,\nonumber\\
     B_2 &=& B^{\rm UV}_2 + {B}_2^{\rm R} , \nonumber\\
     {\delta m}_2 &=& {\delta m}^{\rm UV}_2 . 
\label{L2B2}
\end{eqnarray}
$\delta {m}_2^{\rm R} = 0$ is the specific feature of the {\it K}-operation
for the second-order self-energy diagram.

Substituting Eqs.~(\ref{M4}) and~(\ref{L2B2})
in Eq.~(\ref{a4_standard}) we obtain
\begin{equation}
a_4 = {M}_4^{\rm R} - M_2 ( 2 {L}_2^{\rm R} + {B}_2^{\rm R} ),
\label{a4_intermediate}
\end{equation}
Note that the coefficients of ${L}_2^{\rm R}$ and ${B}_2^{\rm R}$ in
Eq.~(\ref{a4_intermediate}) inherit the coefficients of $L_2$ and $B_2$
in Eq.~(\ref{a4_standard}).


\subsubsection{Separation of IR divergences by the I/R-subtraction}

The second-order mass renormalization is completely carried out and
no remainder is left in the {\it K}-operation.
The {\it R}-subtraction, then,  is not applied by {\sc gencode}{\it N} 
in the case of the fourth order.
IR divergence is caused by a photon spanning over a self-energy-like subdiagram 
which actually represents a lower-order magnetic moment.
This magnetic moment can be effectively represented by a three-point 
vertex between one photon and two  electrons.
Thus,   the UV-finite term ${M}_4^{\rm R}$ 
must have an IR singular structure which is similar to that of the vertex renormalization constant $L_2^{\rm R}$:
\begin{equation}
     M_4^{\rm R} = M_2 L_2^{\rm R}  + \Delta M_4 ,
\label{M4R}
\end{equation}    
where $M_2$ comes from the second-order self-energy subdiagram of $M_{4b}$ and
$L_2^{\rm R}$ appears  by replacing the $M_2$ self-energy subdiagram 
by a point vertex.

The IR-divergence is also found 
in the vertex and wave-function renormalization 
constants.  The WT-identity
\begin{equation}
L_2 + B_2 = 0
\label{ward2}
\end{equation}
guarantees that $L_2$ and $B_2$ have the same, but opposite in sign, IR singularity.
This enables us to separate the IR-singular and  finite terms  of 
$L_2^{\rm R}$ and $B_2^{\rm R}$ as follows:
\begin{eqnarray}
     {L}_2^{\rm R} &=&  I_2 + \Delta{L}_2  ,\nonumber\\
     {B}_2^{\rm R} &=& -I_2 + \Delta{B}_2.
\label{L2B2R}
\end{eqnarray}
where $I_2$ is IR-singular but its finite part is undetermined. 
The finite terms $\Delta L_2$ and $\Delta B_2$ depend on 
how we define $I_2$.
For instance, in Ref.~\cite{Kinoshita:1990}, the {\it I}-operation
was defined so that $I_2 =  L_2^{\rm R} = \ln (\lambda/m) + 5/4$, 
where $\lambda$ is the photon mass.
The sum $L_2^{\rm R} + B_2^{\rm R}$, however, 
does not depend on the definition of $I_2$.
We find that  
\begin{equation}
\Delta \LB_2 \equiv L_2^{\rm R} + B_2^{\rm R} = \Delta L_2 + \Delta B_2 = \frac{3}{4}.
\label{DLB2}
\end{equation}
In other words, the finite quantity $\Delta \LB_2$ is determined 
by how we extract UV divergence by the {\it K}-operation  from each  of $L_2$ and $B_2$:  
\begin{equation}
      L^{\rm UV}_2 + B^{\rm UV}_2 = -\frac{3}{4}. 
\end{equation}

Substituting Eqs.~(\ref{M4R}) and~(\ref{L2B2R}) 
in Eq.~(\ref{a4_intermediate}), one can express $a_4$
defined by the standard renormalization as a sum of finite terms only:
\begin{equation}
a_4 = \Delta {M}_4 - M_2 ~\Delta \LB_2 .
\label{a_4}
\end{equation}
%


\subsection{sixth-order case}

The sixth-order magnetic moment $a_6$ 
has contributions from ten diagrams,
each of which represents the sum of five vertex diagrams transformed with the help of
the WT-identity.
In the standard renormalization it can be written 
in terms of unrenormalized amplitudes $M_6$, $M_4$, etc., 
and various renormalization constants as
\begin{eqnarray}
a_6 &=& M_6 \nonumber \\
    &-& M_4 ( 3 B_2 + 4 L_2) - M_{4^*} {\delta m}_2 \nonumber \\
    &-& M_2 (  B_4 + 2 L_4) - M_{2^*} {\delta m}_4    \nonumber \\
    &+& M_2 \{ 2 (B_2)^2 + 8 B_2 L_2 +7 (L_2)^2 \} 
          + M_{2^*} (3 B_2 + 4 L_2)  {\delta m}_2  \nonumber \\
    &+& M_2 ( B_{2^*} + 4 L_{2^*})  {\delta m}_2  \nonumber \\
    &+& M_{2^{*}} {\delta m}_2 {\delta m}_{2^*} +  M_{2^{**}} ({\delta m}_2 )^2 ,
\label{a6_standard}
\end{eqnarray}
where $M_4$ is defined by Eq.~(\ref{defM4}),
$M_{2^{**}}$ is obtained from $M_2$ by inserting two two-point vertices
in the lepton line of $M_2$, and
\begin{eqnarray}
M_6 &=& \sum_{i=A}^H \eta_i  M_{6i},~~\eta_i =1~~\text{except that}~~ \eta_D = \eta_G = 2, \nonumber \\
M_{4^*} &=&\sum_{i=1}^3 ( M_{4a(i^*)} + M_{4b(i^*)}), \nonumber \\
B_4 &=& B_{4a} + B_{4b}, \nonumber \\
L_4 &=& \sum_{i=1}^3 (L_{4a(i)} + L_{4b(i)}),
\nonumber \\
{\delta m}_4 &=& {\delta m}_{4a} + {\delta m}_{4b}.
\label{defM6andsoon}
\end{eqnarray}
$M_{4a(i^*)}$ is obtained from $M_{4a}$ 
(which contains three lepton lines 1, 2, 3)  by inserting a two-point vertex
in the lepton line $i$ of $M_{4a}$,
and $L_{4a(i)}$ is the vertex renormalization constant of the diagram
in which an external vertex is inserted in the lepton line $i$ of the diagram $4a$.
Similar notation is applied  for the diagrams built from $4b$.

The coefficient of $M_4$ in Eq.~(\ref{a6_standard}) can be readily
understood
noting that the fourth-order self-energy diagrams $M_{4a}$ and $M_{4b}$ have
three fermion lines into which second-order self-energy can be inserted
and four vertices into which second-order vertex can be inserted.
Similarly, there are one fermion line and two vertices
in the second-order self-energy diagram $M_2$
into which we can insert a $B_4$ or a $L_4$, which leads to
$- M_2(B_4 + 2 L_4)$.
The term $M_2 \{ 2 (B_2)^2 + 8 B_2 L_2 + 7 (L_2)^2\}$ comes from
two ways of inserting $B_2$ in $M_2$ (one disjoint and one nested relations
of two $B_2$'s \cite{Aoyama:2005kf}),
eight ways of inserting one $L_2$ and one $B_2$ in $M_2$ (two disjoint, two
overlapping, and four nested relations of $L_2$ and $B_2$),
and seven ways of inserting two $L_2$ in $M_2$ (one disjoint, four
overlapping, and two nested relations of two $L_2$'s).
There is only one way to insert
${\delta m}_4$ in $M_2$
and ${\delta m}_2$ in $B_2$ of $M_2 B_2$.
There are three ways to insert ${\delta m}_2$ in $M_4$, but 
the coefficient three is
included in the definition of $M_{4^*}$.
There are four ways to insert ${\delta m}_2$ in $L_2$ of $M_2 L_2$.
The coefficients of other terms
can be understood in a similar fashion.


\subsubsection{Separation of UV divergences by the {\it K}-operation}

Analysis of the UV divergence structure of $M_6$, 
$L_4$, $B_4$, and ${\delta m}_4$
by the {\it K}-operation leads to
\begin{eqnarray}
 M_6 &=& {M}_6^{\rm R} \nonumber \\ 
     &+& M_4 (3 B^{\rm UV}_2 + 4 L^{\rm UV}_2) + M_{4^*} {\delta m}_2  \nonumber \\
     &+& M_2 (B^{\rm UV}_4 + 2L^{\rm UV}_4) + M_{2^*} {\delta m}^{\rm UV}_4  \nonumber \\
    &-& M_2 \{  (B^{\rm UV}_2)^2 
               + B^{\rm UV}_2 {B}_{2^\prime}^{\rm UV} 
             + 4 B^{\rm UV}_2  L^{\rm UV}_2 
             + 4 B^{\rm UV}_2 {L}_{2^\prime}^{\rm UV} 
             + 7 ( L^{\rm UV}_2)^2  \} \nonumber \\
    &-& M_{2^{*}}  
              (   2  B_2^{\rm UV}  
                 +4  L_2^{\rm UV}  ) {\delta m}_2 \nonumber \\  
    &-& M_{2^{*}}  B_2^{\rm UV}  {\delta m}_{2^\prime}^{\rm UV} \nonumber \\ 
    &-&  M_{2^*}  {\delta m}_{2^*}^{\rm UV} {\delta m}_2  \nonumber \\
    &-&  M_{2^{**}} ({\delta m}_2)^2 ,
\label{M6}
\end{eqnarray}
where
\begin{eqnarray}
{M}_6^{\rm R} &=& \sum_{i=A}^H \eta_i  {M}_{6i}^{\rm R},~~\eta_i =1~~{\rm  except~that}~~ \eta_D = \eta_G = 2, \nonumber \\
\end{eqnarray}
is the UV-finite part of $M_6$.
UV-divergent parts of $L_4$, $B_4$, and ${\delta m}_4$ are separated as follows:
\begin{eqnarray}
 L_4    &=& L^{\rm UV}_4 + 3 L^{\rm UV}_2 L_2^{\rm R} 
         +2 B^{\rm UV}_2 {L}_{2^{'}}^{\rm R} + 2 {\delta m}_2 L_{2^*} + L_4^{\rm R} ,\nonumber\\
 B_4     &=& B_4^{\rm UV} + 2 L^{\rm UV}_2 B_2^{\rm R} 
         +B^{\rm UV}_2 {B}_{2^{'}}^{\rm R} +  {\delta m}_2 B_{2^*} + B_4^{\rm R}, \nonumber \\
 {\delta m}_4    &=& {\delta m}^{\rm UV}_4  + {\delta m}_2 {\delta m}_{2^*}^{\rm R} 
         + B^{\rm UV}_2 {\delta m}_{2^{'}}^{\rm R} + {\delta m}_4^{\rm R}  .
\label{LBdm4}
\end{eqnarray}
$M_4^{\rm R}$ is defined in Eq.~(\ref{M4}), and
$L^{\rm UV}_2$ and  $B^{\rm UV}_2$ are defined in Eq.~(\ref{L2B2}).

Substituting Eqs.~(\ref{M6}), (\ref{M4}), (\ref{LBdm4}), and~(\ref{L2B2})
in Eq.~(\ref{a6_standard}) in this order, we obtain
$a_6$ expressed by UV-finite quantities only:
\begin{eqnarray}
a_6 & = & {M}_6^{\rm R} \nonumber \\
    & - & {M}_4^{\rm R} ( 3 {B}_2^{\rm R} + 4 {L}_2^{\rm R}) \nonumber \\
    & - & M_2 ( {B}_4^{\rm R} + 2 {L}_4^{\rm R})
             -  M_{2^*} {\delta m}_4^{\rm R} \nonumber \\
    & + & M_2 \{  2 ({B}_2^{\rm R})^2 + 8 {B}_2^{\rm R} {L}_2^{\rm R} 
                + 7({L}_2^{\rm R})^2  \}  ~.
\label{a6_intermediate}
\end{eqnarray}
Note that this equation has exactly the same structure as Eq.~(\ref{a6_standard}),
although it looks simpler because ${\delta m}_2^{\rm R} = 0$
in the {\it K}-operation.
This is what one would expect since, in Eq.~(\ref{a6_standard}),
all UV-divergent quantities must cancel out, leaving only UV-finite
pieces with their original numerical coefficients.


\subsubsection{Separation of IR divergences by the I/R-subtraction}

Since Eq.~(\ref{a6_intermediate})
has no linearly IR divergent term caused by the self-mass term,
there is no need to invoke the {\it R}-subtraction.
We, however, retain the {\it R}-subtraction 
that is incorporated in {\sc gencode}{\it N}.
Quantities obtained above can be expressed as the sum of
logarithmically IR-divergent pieces 
defined by the {\it I}-subtraction and finite remainders together with the
residual mass-renormalization term defined by the {\it R}-subtraction:
\begin{eqnarray}
     {M}_6^{\rm R} &=& {L}_4^{\rm R} M_2 - ({L}_2^{\rm R})^2 M_2
                           +{L}_2^{\rm R} {M}_4^{\rm R} + {\delta m}_4^{\rm R} M_{2^*} + \Delta {M}_6 , \nonumber\\
     {L}_4^{\rm R} &=&  I_4 + ({L}_2^{\rm R})^2 + \Delta L_4 \nonumber \\ 
     {B}_4^{\rm R} &=& -I_4 + {L}_2^{\rm R}  {B}_2^{\rm R} +\Delta B_4, 
\label{M6R&LB4R}
\end{eqnarray}
where IR-divergent terms are contained in
$L_2^{\rm R}$, $B_2^{\rm R}$  and  $I_4$ term.
The WT-identity  guarantees that $L_4$ and $B_4$ have 
the same overall IR-divergence which we call $I_4$.
In the previous work \cite{Kinoshita:1990}
the $I_4$ is chosen as the sum of
non-contraction terms $I_{4a(i)}$  of the vertex renormalization 
constants $L_{4a(i)}$:
\begin{equation}
I_4 \equiv  I_{4a(1)} +   I_{4a(2)} + I_{4a(3)} +  I_{4b(1)} +   I_{4b(2)} + I_{4b(3)} .  
\end{equation}
The finite quantities $\Delta L_4$ and $\Delta B_4$ depend on how  $I_4$ is 
defined. But the sum of $L_4^{\rm R} + B_4^{\rm R}$ is independent from the definition of $I_4$. Therefore, we introduce the finite quantity $\Delta \LB_4$ by
\begin{equation}
\Delta \LB_4 \equiv L_4^{\rm R} + B_4^{\rm R} - L_2^{\rm R} \Delta \LB_2 = \Delta L_4 + \Delta B_4 ~.
\label{DLB4}
\end{equation}
Note that the value of
$\Delta \LB_4 $
is unambiguously determined by our choice
of $L^{\rm UV}_4$ and $B^{\rm UV}_4$ in the {\it K}-operation 
and by the WT-identity $L_4 +B_4 =0$.

Substituting Eqs.~(\ref{M6R&LB4R}), (\ref{DLB4}), (\ref{M4R}), and~(\ref{DLB2})
in Eq.~ (\ref{a6_intermediate}) in this order, we obtain
$a_6$ of standard renormalization as the sum of finite terms only
\begin{eqnarray}
a_6 & = & \Delta {M}_6 - 3 \Delta {M}_4 \Delta {\LB}_2 \nonumber \\
    & + & M_2 \{  -\Delta {\LB}_4 + 2 (\Delta {\LB}_2)^2 \}.
\label{a_6}
\end{eqnarray}
%


\subsection{eighth-order case}

The eighth-order magnetic moment $a_8$ 
has contributions from 74 WT-summed diagrams.
In the standard renormalization the renormalized moment $a_8$ can be written 
in terms of unrenormalized amplitudes $M_8$, $M_6$, $M_4$, etc., 
and various renormalization constants as
\begin{eqnarray}
a_8 &=& M_8 \nonumber \\
    &-& M_6 ( 5 B_2 + 6 L_2) \nonumber \\
    &-& M_{6^*} {\delta m}_2 \nonumber \\
    &+& M_4 \{ -3 B_4 - 4 L_4 + 9 (B_2)^2 +26 B_2 L_2 + 18 (L_2)^2 
                + {\delta m}_2 (3 B_{2^*} + 8 L_{2^*} ) \} \nonumber \\
    &+& M_{4^*} \{ {\delta m}_2 (5 B_2 + 6 L_2) +{\delta m}_2 {\delta m}_{2^*} -{\delta m}_4 \} \nonumber \\
    &+& M_{4^{**}} ({\delta m}_2)^2  \nonumber \\
    &+& M_2 \{- B_6 - 2 L_6 +12 L_4 B_2 +18 L_4 L_2 +6 B_4 B_2 + 10 B_4 L_2 \nonumber \\
    & &     -54 B_2 (L_2)^2 -30 (B_2)^2 L_2 - 5 (B_2)^3 -30 (L_2)^3 
                                  \} \nonumber \\
    &+& M_2 {\delta m}_4 (B_{2^*} + 4 L_{2^*} ) \nonumber  \\
    &+& M_2 {\delta m}_2 ( B_{4^*} + 2 L_{4^*} -6 B_2 B_{2^*} -24 B_2 L_{2^*} -10 B_{2^*} L_2 -36 L_2 L_{2^*} ) \nonumber  \\
    &-& M_2 {\delta m}_2 {\delta m}_{2^*} (  B_{2^*} +4 L_{2^*} ) \nonumber  \\
    &-& M_2 ({\delta m}_2)^2  (  B_{2^{**}} +4 L_{2^{**\dagger}} +2 L_{2^{*\dagger *}} ) \nonumber  \\
    &+& M_{2^*} {\delta m}_2 \{ 3 B_4 +4  L_4 + {\delta m}_{4^*} -26 B_2 L_2 -9 (B_2)^2 -18 (L_2)^2  \} \nonumber  \\
    &-& M_{2^*} {\delta m}_6   \nonumber  \\
    &+& M_{2^*} {\delta m}_4 (5 B_2 + 6 L_2 + {\delta m}_{2^*} ) \nonumber  \\
    &-& M_{2^*} ({\delta m}_2)^2  ( 3  B_{2^*} +8 L_{2^*} +{\delta m}_{2^{**}} ) \nonumber  \\
    &-& M_{2^*} {\delta m}_2 {\delta m}_{2^*} ( 5 B_2 +6 L_2 ) \nonumber  \\
    &-& M_{2^*} {\delta m}_2 ({\delta m}_{2^*})^2  \nonumber  \\
    &+& M_{2^{**}} {\delta m}_2  \{2 {\delta m}_4  - {\delta m}_2 (  5 B_2 +6 L_2  +2 {\delta m}_{2^*}) \}  \nonumber  \\
    &-& M_{2^{***}} ({\delta m}_2)^3 .
\label{a8_standard}
\end{eqnarray}
$M_8$ is defined by
\begin{equation}
M_8 = \sum_{\alpha=01}^{47} \eta_\alpha  M_{\alpha},
\end{equation}
where
$\eta_\alpha$ = 1 for time-reversal-symmetric diagrams
and 
$\eta_\alpha$ = 2 for others.


\subsubsection{Separation of UV divergences by the {\it K}-operation}

The UV divergence structure of $M_8$ is given by
\begin{eqnarray}
M_8 &=& M_8^{\rm R}
\nonumber \\
       &+& M_6   ( 5 B_2^{\rm UV} + 6 L_2^{\rm UV} )
\nonumber \\
      &+& M_{6^*}    {\delta m}_2 
\nonumber \\
      &+& M_4  \{ 3 B_4^{\rm UV} + 4 L_4^{\rm UV} 
         - 3 B_2^{\rm UV} B_{2^\prime}^{\rm UV} 
         - 6 (B_2^{\rm UV})^2 
         - 18 B_2^{\rm UV} L_2^{\rm UV} 
         - 8 B_2^{\rm UV} L_{2^\prime}^{\rm UV}
         - 18 (L_2^{\rm UV})^2
                       \} 
\nonumber \\
    &+& M_{4^*}   ( {\delta m}_4^{\rm UV} - B_2^{\rm UV} {\delta m}_{2^\prime}^{\rm UV} 
           - 4 {\delta m}_2 B_2^{\rm UV} - 6 {\delta m}_2 L_2^{\rm UV} 
           - {\delta m}_2 {\delta m}_{2^*}^{\rm UV} )
\nonumber \\
    &-& M_{4^{**}}   ({\delta m}_2)^2 
\nonumber \\
 &+& M_2   \{ B_6^{\rm UV} + 2 L_6^{\rm UV} 
       - 2 B_4^{\rm UV} B_2^{\rm UV} - 6 B_4^{\rm UV} L_2^{\rm UV} 
       - B_4^{\rm UV} B_{2^\prime}^{\rm UV} 
       - 4 B_4^{\rm UV} L_{2^\prime}^{\rm UV} 
\nonumber \\
 & &   - 4 L_4^{\rm UV} B_2^{\rm UV} - 18 L_4^{\rm UV} L_2^{\rm UV} 
       + 6 B_2^{\rm UV} L_2^{\rm UV} B_{2^\prime}^{\rm UV} 
       + 36 B_2^{\rm UV} L_2^{\rm UV} L_{2^\prime}^{\rm UV} 
       + 18 B_2^{\rm UV} (L_2^{\rm UV})^2 
\nonumber \\
  &&   - B_2^{\rm UV} B_{4^\prime}^{\rm UV} 
       - 2 B_2^{\rm UV} L_{4^\prime}^{\rm UV}
       + 4 B_2^{\rm UV} B_{2^\prime}^{\rm UV} L_{2^\prime}^{\rm UV} 
       + B_2^{\rm UV} (B_{2^\prime}^{\rm UV})^2 
       + 6 (B_2^{\rm UV})^2 L_2^{\rm UV} 
\nonumber \\
  &&   + 2 (B_2^{\rm UV})^2 L_{2^{\prime \prime}}^{\rm UV} 
       + (B_2^{\rm UV})^2 B_{2^{\prime \prime}}^{\rm UV} 
       + 2 (B_2^{\rm UV})^2 B_{2^\prime}^{\rm UV} 
\nonumber \\
  &&   + 8 (B_2^{\rm UV})^2 L_{2^\prime}^{\rm UV} 
       + (B_2^{\rm UV})^3 + 30 (L_2^{\rm UV})^3  \} 
\nonumber \\
 & + &  M_{2^*} \{ {\delta m}_6^{\rm UV} 
             - B_2^{\rm UV} {\delta m}_{4^\prime}^{\rm UV} 
             + (B_2^{\rm UV})^2 {\delta m}_{2^{\prime \prime}}^{\rm UV} 
                \}
\nonumber \\
 & + &  M_{2^*}  {\delta m}_4^{\rm UV}  \{
             - 2  B_2^{\rm UV} - 6 L_2^{\rm UV} 
             - {\delta m}_{2^*}^{\rm UV} 
             \}
\nonumber \\
 & + &  M_{2^*} {\delta m}_{2^\prime}^{\rm UV} \{
             - B_4^{\rm UV} 
             + 6 B_2^{\rm UV} L_2^{\rm UV} 
             + 2 (B_2^{\rm UV})^2 
             + B_2^{\rm UV} B_{2^\prime}^{\rm UV} 
             + B_2^{\rm UV} {\delta m}_{2^*}^{\rm UV} 
            \}
\nonumber \\
 & + &  M_{2^*} {\delta m}_2\{
             - 2 B_4^{\rm UV} - 4  L_4^{\rm UV} -  {\delta m}_{4^*}^{\rm UV} 
             + 18  (L_2^{\rm UV})^2
\nonumber \\
       &&    +  B_2^{\rm UV} (12 L_2^{\rm UV} + 8 L_{2^\prime}^{\rm UV} 
             +  2 B_{2^\prime}^{\rm UV} + 3  B_2^{\rm UV}
             + 2 {\delta m}_{2^{*\prime}}^{\rm UV}) 
\nonumber \\
        &&   + {\delta m}_{2^*}^{\rm UV}  ( 2 B_2^{\rm UV} +  6  L_2^{\rm UV} 
                                   +  {\delta m}_{2^*}^{\rm UV} )
             \}
\nonumber \\
 &+& M_{2^{**}}  {\delta m}_2 \{ -2 {\delta m}_4^{\rm UV} 
                    + (3 {\delta m}_2  
                    + 2 {\delta m}_{2^\prime}^{\rm UV} )B_2^{\rm UV} 
                    + 6 {\delta m}_2 L_2^{\rm UV} 
                    + 2 {\delta m}_2 {\delta m}_{2^*}^{\rm UV} \} 
\nonumber \\
 &+& M_{2^{***}}  ( {\delta m}_2)^3 ~.
\label{M8}
\end{eqnarray}

The quantities with a prime, $L_{2^\prime}$, called derivative amplitudes,
arises from a fourth-order diagram that contains a self-energy subdiagram.
This self-energy subdiagram supplies the inverse of the fermion propagator
times the wave function renormalization constant and cancels out one of 
the adjacent fermion propagators and yields another renormalization constant $L_2$.
In the expression $L_{2^\prime}$,  the inverse fermion propagator and the adjacent fermion
propagator are left in the numerator and denominator, respectively,
of the Feynman parametric expression of the amplitude.
Thus the whole renormalization constant $L_{2^\prime}$ is analytically 
identical with $L_2$. But, the separation of UV divergence 
by the {\it K}-operation works differently in two cases, 
so that $L_{2^\prime}^{\rm UV}$ is different from $L_2^{\rm UV}$ 
by a finite amount.

A similar consideration applies to higher order quantities.
Take, for instance, $L_4$ which consists of four fermion lines. 
There are four ways to insert a self-energy subdiagram to $L_4$. 
Since we define $L_{4^\prime}$ as the sum of all
derivative amplitudes of $L_4$, we have $ L_{4^\prime} = 4 L_4$.  
Similarly, $B_{4^\prime} = 3 B_4$.

The second-order derivative amplitude, such as $L_{2^\prime}$, however, does not include its symmetric factor in our definition. Thus, $L_{2^\prime}=L_2$.   

We also need the UV divergence structures of the renormalization terms $B_6$, 
$L_6$, and  ${\delta m}_6$:
\begin{eqnarray}
     B_{6} &=&   B_{6}^{\rm UV}  + B_{6}^{\rm R}
\nonumber \\
       &+& B_{4^*}  {\delta m}_2 
\nonumber \\
       &+& B_2^{\rm R} \{ 2 L_{4}^{\rm UV} - 4 B_2^{\rm UV} L_{2^{\prime}}^{\rm UV} - 7 (L_{2}^{\rm UV})^2 \} 
\nonumber \\
       &+& B_{2^{\prime}}^{\rm R} ( B_{4}^{\rm UV} - 4 B_2^{\rm UV} L_{2}^{\rm UV} - B_2^{\rm UV} B_{2^{\prime}}^{\rm UV} )
\nonumber \\
       &-& B_{2^{\prime \prime}}^{\rm R} (   B_2^{\rm UV})^2 
\nonumber \\
       &+& B_{2^{* \prime}}^{\rm R} (  - 2 {\delta m}_2 B_2^{\rm UV} )
\nonumber \\
       &+& \widetilde{B_{4}} ( 4 L_{2}^{\rm UV} )
\nonumber \\
       &+& \widetilde{B_{4^{\prime}}} ( B_2^{\rm UV} )
\nonumber \\
       &+& B_{2^*} ( {\delta m}_4^{\rm UV} - B_2^{\rm UV} {\delta m}_{2^{\prime}}^{\rm UV} - 4 {\delta m}_2 L_{2}^{\rm UV} - {\delta m}_2 {\delta m}_{2^*}^{\rm UV} )
\nonumber \\
       &-& B_{2^{**}} ({\delta m}_2)^2,
\label{B6}
\end{eqnarray}
\begin{eqnarray}
   L_6 &= &L_6^{\rm UV}  + L_6^{\rm R}
\nonumber \\
       &+ & L_{4^*} ( {\delta m}_2 )
\nonumber \\
       & +&  L_2^{\rm R} \{ 3 L_{4}^{\rm UV} - 6 B_2^{\rm UV} L_{2^{\prime}}^{\rm UV} 
       - 12 (L_2^{\rm UV})^2 \} 
\nonumber \\
       & + &  L_{2^{\prime}}^{\rm R} ( 2 B_{4}^{\rm UV} - 10 B_2^{\rm UV} L_2^{\rm UV} - 2 B_2^{\rm UV} B_{2^{\prime}}^{\rm UV} )
\nonumber \\
       & - &  L_{2^{\prime \prime}}^{\rm R} (   B_2^{\rm UV})^2 
\nonumber \\
       & + &  \widetilde{L_{4}} ( 5 L_2^{\rm UV} )
\nonumber \\
       & + & \widetilde{L_{4^{\prime}}} ( B_2^{\rm UV} )
\nonumber \\
       & + & L_{2^{*\prime}} ^{\rm R}(  - 2 {\delta m}_2 B_2^{\rm UV} )
\nonumber \\
       & + & L_{2^*} ( 2 {\delta m}_4^{\rm UV} - 2 B_2^{\rm UV} {\delta m}_{2^{\prime}}^{\rm UV} - 10 {\delta m}_2 L_2^{\rm UV} - 2 {\delta m}_2 {\delta m}_{2^*}^{\rm UV} )
\nonumber \\
       &- &L_{2^{**}} (   {\delta m}_2)^2 ~,
\label{L6}
\end{eqnarray}
and
\begin{eqnarray}
 {\delta m}_{6}&=& {\delta m}_{6}^{\rm UV} + {\delta m}_{6}^{\rm R}
\nonumber \\
       &-& {\delta m}_{2^{\prime \prime}}^{\rm R}  (  B_2^{\rm UV})^2 
\nonumber \\
       &+& {\delta m}_{2^\prime}^{\rm R}  
            ( B_4^{\rm UV} - B_2^{\rm UV} B_{2^\prime}^{\rm UV} )
\nonumber \\
       &+& {\delta m}_{2^*}^{\rm R}  ( {\delta m}_4^{\rm UV} - B_2^{\rm UV} {\delta m}_{2^\prime}^{\rm UV} 
              - {\delta m}_2 {\delta m}_{2^*}^{\rm UV} )
\nonumber \\
       &+& {\delta m}_{2^{* \prime}}^{\rm R}   (  - 2 {\delta m}_2 B_2^{\rm UV} )
\nonumber \\
       &+& {\delta m}_4^{\rm R}  ( 4 L_2^{\rm UV} )
\nonumber \\
       &+& \widetilde{{\delta m}_{4^*}}   {\delta m}_2 
\nonumber \\
       &+& \widetilde{{\delta m}_{4^\prime}}  B_2^{\rm UV} 
\nonumber \\
       &-& {\delta m}_{2^{**}}   ({\delta m}_2)^2 ,
\label{dm6}
\end{eqnarray}
where the quantity $\widetilde{A}$ 
is defined by $\widetilde{A} \equiv A - A^{\rm UV}$. 
The difference between $\widetilde{A}$ and $A^{\rm R}$ is 
that  the former contains 
UV divergent terms arising from subdiagrams, while the latter
is completely free from these sub-UV divergences.
For instance, we have 
\begin{eqnarray}
   \widetilde{B_4}&\equiv& B_4- B_4^{\rm UV}
\nonumber \\
                  &=&  B_4^{\rm R} + {\delta m}_2 B_{2^*} + B_2^{\rm UV} B_{2^{\prime}}^{\rm R} 
                                  + 2 L_2^{\rm UV} B_2^{\rm R},
 \\
   \widetilde{L_4}&\equiv & L_4- L_4^{\rm UV}
\nonumber \\
                  &=&  L_4^{\rm R} + 2 {\delta m}_2 L_{2^*} + 2 B_2^{\rm UV} L_{2^{\prime}}^{\rm R} 
                                  + 2 L_2^{\rm UV} L_2^{\rm R},
\end{eqnarray}
and so on.

We also need the UV divergence structure of 
$M_{4^*}$, which is 
the amplitude of the fourth-order magnetic moment with a two-point 
vertex insertion: 
\begin{equation}
M_{4^*} = M_{4^*}^{\rm R} + 2 L_2^{\rm UV} M_{2^*} 
        + 2 ( {\delta m}_2 M_{2^{**}} + B_2^{\rm UV} M_{2^{*}} ) 
        + {\delta m}_{2^*}^{UV} M_{2^*} ~.
\label{M4s}
\end{equation}

Substituting the UV structures of the eighth order Eq.~(\ref{M8}), the sixth-order quantities Eqs.~(\ref{M6}), (\ref{B6}), (\ref{L6}) and~(\ref{dm6}), those of the fourth order 
Eqs.~(\ref{M4}), (\ref{LBdm4}), and~(\ref{M4s}), and those of the second order (\ref{L2B2}) in this sequence  
in Eq.~(\ref{a8_standard}),  we obtain
the UV-finite expression of the magnetic moment $a_8$:
\begin{eqnarray}
a_8 & = & {M}_8^{\rm R} \nonumber \\
 & + & {M}_6^{\rm R} (-5{B}_2^{\rm R} -6{L}_2^{\rm R} ) \nonumber \\
 & + & {M}_4^{\rm R} \{  -4 {L}_4^{\rm R} -3 {B}_4^{\rm R} 
                         +26 {L}_2^{\rm R} {B}_2^{\rm R} 
                         + 18 ({L}_2^{\rm R})^2 + 9 ({B}_2^{\rm R})^2 
                      \}  \nonumber \\
 & - & {M}_{4^*}^{\rm R} {\delta m}_4^{\rm R} \nonumber \\
 & + & M_2 \{ 
      - {B}_6^{\rm R} -2 {L}_6^{\rm R} 
      + ({B}_{2^*} +4 {L}_{2^*}) {\delta m}_4^{\rm R}    \nonumber \\
    &&+ 6 {B}_4^{\rm R} {B}_2^{\rm R}
      + 10 {B}_4^{\rm R} {L}_2 ^{\rm R}
      + 12 {B}_2^{\rm R} {L}_4^{\rm R} 
      + 18 {L}_2^{\rm R} {L}_4^{\rm R} \nonumber \\
    &&- 30 {L}_2^{\rm R} ({B}_2^{\rm R})^2 
      - 54 ({L}_2^{\rm R})^2 {B}_2^{\rm R}
      -30 ({L}_2^{\rm R})^3 
      -5 ({B}_2^{\rm R})^3 
            \} 
\nonumber \\
    & + & M_{2^*}\{ - {\delta m}_6^{\rm R} 
+ {\delta m}_4^{\rm R} ({\delta m}_{2^*}^{\rm R}
          +6 {L}_2^{\rm R} + 5 {B}_2^{\rm R} )\}~.
\label{a8_intermediate}
\end{eqnarray}
Again Eq.~(\ref{a8_intermediate}) has exactly the same structure as
Eq.~(\ref{a8_standard}) except that ${\delta m}_2^{\rm R} = 0$.


\subsubsection{I/R-subtraction }

In order to handle the numerical calculation  on a computer, we need
to separate the IR divergent terms from $M_8^{\rm R}$.
Paying attention to the outermost photon  spanning  over a self-energy
subdiagram,   we obtain the  IR structure of $M_8^{R}$ as follows: 
\begin{eqnarray}
M_8^{\rm R} 
        &=& \Delta M_8 \nonumber \\
        &+& M_6^{\rm R} L_2^{\rm R}  \nonumber \\
        &+& M_4^{\rm R} \{  L_4^{\rm R} -(L_2^{\rm R})^2  \} \nonumber \\
        &+& M_2 \{ L_6^{\rm R} - 2 L_4^{\rm R} L_2^{\rm R} + (L_2^{\rm R})^3 
              -2 {\delta m}_4^{\rm R} L_{2^*}^{\rm R} \}\nonumber \\
        &+& M_{4^*}^{\rm R} {\delta m}_4^{\rm R}  
\nonumber \\
        &+& M_{2^*} ( 
        {\delta m}_6^{\rm R} - {\delta m}_4^{\rm R} {\delta m}_{2^*}^{\rm R} -{\delta m}_4^{\rm R} L_2^{\rm R} )~.
\label{M8R}
\end{eqnarray}
Eq.~(\ref{M8R}) has a term
${M}_{4^*}^{\rm R} {\delta m}_4^{\rm R}$, 
where $M_{4^*}^{\rm R}$ is linearly IR-divergent, which arises from 
the diagrams $M_{16}$ and $M_{18}$.
This term compensates the same IR-divergence found 
in $a_8$ of Eq.~(\ref{a8_intermediate}) whose origin is the mass-renormalization 
term $M_{4^*} {\delta m}_4$ associated with  the diagrams  $M_{16}$ and $M_{18}$.
This IR divergence in $M_8^{\rm R}$ can thus be removed 
from $M_{16}$  and $M_{18}$
by the {\it R}-subtraction which acts  on a fourth-order self-energy 
subdiagram of $M_{16}$ ($M_{18}$)
and  complements the mass-renormalization constant ${\delta m}_{4a}({\delta m}_{4b})$.

Another linear IR divergent term in Eq.~(\ref{M8R})  is 
$2 M_2 L_{2^*}^{\rm R} {\delta m}_4^{\rm R}$ , where $L_{2^*}$ is linearly divergent.
This IR divergence is again found  in the diagrams of $M_{16}$ and $M_{18}$.
In the IR-limit of the outermost photon loop, 
a possible configuration of $M_{16}(M_{18})$ is that  
the second-order self-energy subdiagram of $M_{16}(M_{18})$
supplies the second-order anomalous magnetic moment $M_2$ and
the fourth-order self-energy subdiagram behaves as  ${\delta m}_{4b(a)}^{\rm R}$.
The IR behavior of the residual diagram including the outermost photon line 
resembles the second-order vertex diagram with a
two-point vertex insertion $L_{2^*}^{\rm R}$:
\begin{equation}
L_{2^*} \equiv \Delta L_{2^*} + L_{2^*}^{\rm R} ~,
\end{equation}
where $\Delta L_{2^*}=-3/4$ is the one contraction term of $L_{2^*}$
and the IR divergent $L_{2^*}^{\rm R}$  is  the non-contraction term of $L_{2^*}$.

This IR divergence in $M_{8}^{\rm R}$ of Eq.~(\ref{M8R}) compensates 
the IR divergence in  $2 M_2 L_{2^*} {\delta m}_4 $  
of the renormalized magnetic moment $a_8$ of Eq.~(\ref{a8_intermediate}).
The origin of the $+4 L_{2^*} {\delta m}_4 M_2$  in Eq.~(\ref{a8_intermediate}) is the renormalization terms
associated with the diagrams $M_{08}$, $M_{10}$, $M_{41}$, and $M_{46}$.
Two of four $L_{2^*}$ terms are exactly canceled by $B_{2^*}$ terms
because of the WT-identity $2 L_{2^*} +  B_{2^*}=0$.
The remaining two $L_{2^*}$ terms will cancel the IR-divergence
arising from the diagrams $M_{16}$ and $M_{18}$ in $M_8^{\rm R}$. 

The last of the linearly IR divergent terms of $M_8^{\rm R}$  
of Eq.~(\ref{M8R}) is $M_2 L_6^{\rm R}$, which also comes from
$M_{16}$ and $M_{18}$. In this case, the second-order self-energy
subdiagram supplies a second-order anomalous magnetic moment $M_2$ and the rest
of the residual diagrams are pushed in the IR limit. From  
$M_{16} (M_{18})$,
it gives rise to the similar IR behavior of 
the six-order vertex renormalization constant
$L_{6b(1)}(L_{6c(1)})$.  This IR divergence will be canceled in $a_8$ of
Eq.~(\ref{a8_intermediate}) by the $M_2 L_6^{\rm R}$ term which
comes from the renormalization constants for the diagram $M_{08}(M_{10})$.

To see the cancellation of remaining logarithmic IR divergence in $a_8$, we need the
IR  structures of the renormalization constants $L_6^{\rm R}$ and $B_6^{\rm R}$.
Resorting to the WT-identity again,  we can define the finite quantity
$\Delta \LB_6$ as follows:
\begin{eqnarray}
\Delta \LB_6 &\equiv & L_6^{\rm R} + B_6^{\rm R} \nonumber \\
              & - & 
                \{ +  I_6 
                  +  2 L_4^{\rm R} L_2^{\rm R} 
                  -    (L_2^{\rm R})^3
                  +  2 {\delta m}_4^{\rm R} L_{2^*}
                \} 
\nonumber \\
              & - & 
                 \{-   I_6 
                 +    L_4^{\rm R} B_2^{\rm R} 
                 +    L_2^{\rm R} B_4^{\rm R} 
                 -    (L_2^{\rm R})^2 B_2^{\rm R}
                 +    {\delta m}_4^{\rm R} B_{2^*} 
                \}~,
\label{DLB6} 
\end{eqnarray}
where $I_6$ is the overall IR divergent term of $L_6$ and $B_6$.
The WT-identity guarantees that $\Delta \LB_6$ 
is independent of the choice of $I_6$.
Note that $\Delta \LB_6 \equiv \Delta L_6 + \Delta B_6 + \Delta L_4 \Delta B_2
+ \Delta {\delta m}_4 B_{2^{*}} [I]$, where the quantities in the right-hand side
are defined in Ref.~\cite{Kinoshita:1990}.


\subsubsection{Finite expression}

Separating  the UV-finite quantities in $a_8$ of Eq.~(\ref{a8_intermediate}) 
into the IR-singular parts and the finite parts as given in Eqs.~(\ref{M8R}),
(\ref{M6R&LB4R}), (\ref{M4R}),  (\ref{DLB6}), (\ref{DLB4}), and~(\ref{DLB2}), 
we obtain the expression $a_8$ in terms of the finite quantities only:
\begin{eqnarray}
a_8 & = & \Delta{M}_8 \nonumber \\
    & + & \Delta{M}_6 (-5  \Delta \LB_2) \nonumber \\
    & + & \Delta{M}_4 \{ -3 \Delta \LB_4 + 9 (\Delta \LB_2)^2  \} \nonumber \\
    & + & M_2 \{ -\Delta \LB_6 +  6 \Delta \LB_4 \Delta \LB_2 
               - 5 (\Delta \LB_2)^3 \} \nonumber \\
    & + & 2 M_2 \Delta L_{2^*}   \Delta {\delta m}_4 . 
\label{a_8}
\end{eqnarray}
Since $\Delta \LB_4 = \Delta L_4 + \Delta B_4$, $2 \Delta L_{2^*} = -\Delta B_{2^*}$,  and $\Delta \LB_2 = \Delta B_2$, 
this is equivalent to Eq.~(76) of Ref.~\cite{Aoyama:2007mn}, which was obtained from the direct sum of all subtraction terms.
Note that the last term  of Eq.~(\ref{a_8}) remains 
unsubtracted regardless of the {\it R}-subtraction,
which is the residual mass-renormalization.
This is because we
use only the non-contraction term $L_2^{\rm R}$ as the IR-subtraction term,
leaving the finite part of $\Delta L_{2^*}$ untouched. 

The definition of the finite term $\Delta L_{2^*}$ does depend on
how to separate IR part from $L_{2^*}$. To avoid such arbitrariness, we stick to the
same {\it I}-subtraction rule of IR separation which is used for vertex renormalization constants.
Namely, the IR-singularity is confined in $L_n^{\rm R}$, which is 
defined by the rule
\begin{eqnarray}
\widetilde{L_n}&\equiv & L_n - \text{highest contraction term of $L_n$} 
\nonumber \\
L_n^{\rm R}    &\equiv & \widetilde{L_n}  - \text{ UV sub-divergence term determined  by {\it K}-operation on  
                                                    $\widetilde{L_n}$}  ~
\end{eqnarray}
and this $L_n^{\rm R}$ is used as an IR-subtraction term.
This determines $\Delta L_{2^*} = - 3/4$ unambiguously.
The {\it K}-operation  does not pick up  this $\Delta L_{2^*}$ term from 
a corresponding subdiagram, since $L_{2^*}$ is UV-finite.
So, no rule exists in the automation code {\sc gencode}{\it N} that allows us
to subtract the finite term  $\Delta L_{2^*}$ of a renormalization constant.

The residual renormalization scheme for the Set IV contribution
$A_1^{(10)} [\text{Set IV}^{(l_1l_2)}]$ can be readily obtained from Eq.~(\ref{a_8})
by insertion of a closed loop of the lepton $l_2$ in the internal
photon lines of $a_8$. This leads to Eq.~(\ref{a8_setIV})
given in Sec.~\ref{sec:residual}.